\documentclass[11pt,a4paper]{article} 
\usepackage{style}

\usepackage{amsmath,amssymb,epsfig,bm,amsfonts,yfonts,bbm,mathrsfs}
\usepackage{slashed}

\title{Real Clifford algebras and their spinors for relativistic fermions}

\author{Stefan Floerchinger}
\emailAdd{stefan.floerchinger@thphys.uni-heidelberg.de}
\affiliation{Institut f\"{u}r Theoretische Physik, Universit\"{a}t Heidelberg, Philosophenweg 16, 69120 Heidelberg, Germany}

\abstract{Real Clifford algebras for arbitrary number of space and time dimensions as well as their representations in terms of spinors are reviewed and discussed. The Clifford algebras are classified in terms of isomorphic matrix algebras of real, complex or quaternionic type. Spinors are defined as elements of minimal or quasi-minimal left ideals within the Clifford algebra and as representations of the pin and spin groups. Two types of Dirac adjoint spinors are introduced carefully. The relation between mathematical structures and applications to describe relativistic fermions is emphasized throughout.}

\begin{document}
\maketitle

\section{Introduction}

Clifford algebra plays an important role for understanding physical theories of relativistic fermions. In the physics literature they are typically discussed for a given number of dimensions in terms of a concrete matrix realization \cite{ZinnJustin:1989mi, Weinberg:1995mt, Wetterich:1982ed, VanProeyen:1999ni,Todorov:2011dr}. In this formulation, spinors are in general complex although additional ``reality constraints'' are sometimes imposed, e.\ g.\ to describe Majorana fermions in $d=1+3$ dimensions. One also has to specify additional relations for time and space reversal transformations as well as different conjugations in spinor space such as charge conjugation. While this approach is well suited for many applications, it is not as systematic as one may wish. A matrix representation must be developed for every combination of space and time dimensions and finding the right realization of discrete symmetries is a kind of art. Other expositions focus more on geometry, but are also restricted to four dimensions \cite{PenroseRindler}.

On the other side, in the mathematical literature, a more systematic approach to Clifford algebras has been developed, see e.\ g.\ refs.\ \cite{Chevalley, Cartan, Lawson:1998yr, ReeseHarvey, Porteous, PlymenRobinson, GilbertMurray, Lounesto:2001zz, Garling:2011zz, Vaz:2016qyw}. This allows to understand and classify real Clifford algebras for $r$ time and $d-r$ space dimensions in rather general terms. Besides the spin group which is so important to physics, also representations of the larger ``pin'' group are being discussed. Because the latter is a double cover of the full indefinite orthogonal group O$(r,d-r,\mathbbm{R})$, it includes also reflections of single coordinates and therefore time and space reversal.  

With the present article we aim at bridging between the mathematical literature and literature about applications in physics, specifically for relativistic fermions. We will develop the theory of real Clifford algebras with indefinite but non-degenerate symmetric metric in a systematic but concise way and provide the material to describe spinor representations for relativistic fermions. The present paper concentrates entirely on {\it real} Clifford algebras, i.\ e.\ the algebra that arises if only the original generators, their various products and linear superpositions thereof with {\it real} coefficients are allowed. The complexified version of this (with complex linear superpositions allowed) is actually of a somewhat simpler structure and more often discussed in the physics literature. Restrictions can then be imposed in addition, specifically in the form of real or quaternionic structures on a complex vector space, in order to describe (variations of) Majorana fermions. We plan to address complex Clifford algebras in a future study but find it natural to start with the more restricted real case.

For more clarity, in the present work we will follow the mathematical tradition to separate the Clifford algebra somewhat from it's concrete matrix representation. There is, however, in any case a rather close connection, because matrix algebras serve not only as representations but can also be used to classify Clifford algebras. We will also make the effort so discuss spinors somewhat independent of the matrix representation. Algebraically, spinor spaces can in fact be introduced as (minimal or quasi-minimal) left and right ideals {\it within} the Clifford algebra \cite{LounestoWene}. Even though it will be some effort to introduce the necessary algebraic notions, this is an effort that pays off eventually because it leads to more clarity on the classification of spinor spaces. Specifically, one finds spinor spaces as real vector spaces, as complex vector spaces, or as quaternionic vector spaces. In particular the quaternionic spaces come somewhat as a surprise from a physicists point of view but arise unavoidably for certain combinations of $d$ and $r$, as will become clear.

For a consistent description of spinor spaces for arbitrary number of time and space dimensions, a rather important role is played by the so-called Clifford structure map $\varsigma$. For even number of dimensions $d$, this structure map is just an element of the Clifford algebra itself and can be taken to be the product of all time-like and space-like generators. For odd number of dimensions $d$, the situations is more complex and the structure map is not an element of the Clifford algebra itself. As we will point out, the structure map plays an important role in defining a consistent action of the pin group on arbitrary elements of the Clifford algebra and in particular on spinor representations as left and right ideals. While the present paper is to a large extend of review character, it is mainly in this context that some discussions given here go beyond the previously available literature.

\section{Real, indefinite orthogonal groups}
\label{sec:RealIndefiniteOrthogonalGroups}

We concentrate on {\it real} coordinates $x^\mu$ and consider $d$ spacetime dimensions divided into $r$ time and $(d-r)$ space dimensions. Technically, we mean by this that there is a real, indefinite but non-degenerate metric, which by a convenient choice of coordinates can be brought to the form $\eta_{\mu\nu}=\text{diag}(-1,\ldots,-1,+1,\ldots,+1)$. The first $r$ entries are $-1$ for time-like coordinates followed by $(d-r)$ entries $+1$ for space coordinates. We label the indices $\mu$ of coordinates $x^\mu$ and the metric such that $\mu=1-r, \ldots, 0$ are time indices and $\mu=1,\ldots, d-r$ are spatial indices.\footnote{We are using here conventions such that the Minkowski space metric has signature $(-,+,+,+)$. One may alternatively use conventions where time coordinates have positive space coordinates negative entries in the metric.}

The spacetime symmetry group (including rotations, boosts and reflections but without translations) is then the one of the indefinite orthogonal group O$(r,d-r, \mathbbm{R})$. The group elements of O$(r,d-r, \mathbbm{R})$ are real $d\times d$ matrices $\Lambda^\mu_{\;\;\nu}$, defined through the relation
\begin{equation}
\Lambda^\rho_{\;\;\mu}\eta_{\rho\sigma} \Lambda^\sigma_{\;\;\nu} = \eta_{\mu\nu}, \quad\quad\quad\text{or}\quad\quad\quad \Lambda^T \eta \Lambda = \eta.
\label{eq:GeneralizedLorentzGroup}
\end{equation}
In other words, these transformations are such that the metric is left invariant. 

\paragraph{Rotation group O$(d, \mathbbm{R})$.}
Let us first discuss the simplest and definite case $r=0$ (or, essentially equivalent, $d=r$). In this case there are two disconnected components of the group O$(d, \mathbbm{R})$ with $\det(\Lambda)=\pm 1$. The elements close to the unit element $\Lambda=\mathbbm{1}$ have $\det(\Lambda)=1$ and form the Lie group SO$(d, \mathbbm{R})$. They can be combined with reflections to construct other elements of the group O$(d, \mathbbm{R})$. For $d$ odd one has a full reflection $\Lambda=-\mathbbm{1}$ with $\det(\Lambda)=-1$. This transformation commutes with all other elements. One has therefore the structure $\text{O}(d, \mathbbm{R})=Z_2 \times \text{SO}(d, \mathbbm{R})$. For $d$ even this is not possible, and reflections with $\det(\Lambda)=-1$ do not commute with all elements of SO$(d)$. In any case, the topology of O$(d, \mathbbm{R})$ has two disconnected parts with $\det(\Lambda)=\pm 1$. 
\begin{table}
\centering
\begin{tabular}{ c | c | c }
\hline\hline
 & $\det(Q)=+1$ & $\det(Q)=-1$ \\ \hline
 $ \det(P)=+1$ & 
 \begin{tabular}{@{}c@{}} I \\ $\det(\Lambda) = +1$ \end{tabular}
 & 
 \begin{tabular}{@{}c@{}} II \\ $\det(\Lambda) = -1$ \end{tabular}
 \\ \hline
$\det(P)=-1$ & 
\begin{tabular}{@{}c@{}} III \\ $\det(\Lambda) = -1$ \end{tabular}
& 
 \begin{tabular}{@{}c@{}} IV \\ $\det(\Lambda) = +1$ \end{tabular} \\
\hline\hline
\end{tabular}
\caption{Topologically disconnected components I, II, III, and IV of the indefinite orthogonal group O$(r,d-r, \mathbbm{R})$. Only subsector I constitutes a Lie group. The other sectors can be written as elements of I combined with reflections $\Lambda=\text{diag}(P,Q)$, as described in the text. In the complexified group O$(r,d-r, \mathbbm{C})$ region I and IV are connected, as well as II and III, but the the two sectors with $\text{det}(\Lambda)=\pm 1$ remain disconnected from each other.}
\label{tab:TopologyLorentzGroup}
\end{table}

\paragraph{Generalized Lorentz group O$(r,d-r,\mathbbm{R})$.} Now we assume $r>0$ and $(d-r)>0$. Again there are two disconnected parts with $\det(\Lambda)=\pm1$. The elements that are smoothly connected to $\Lambda=\mathbbm{1}$ have $\det(\Lambda)=1$. Reflections along the coordinate axis can be written as $\Lambda=\text{diag}(P,Q)$ and can have $\det(P)=\pm 1$, $\det(Q)=\pm 1$ with $\det(\Lambda)=\det(P) \det(Q)$. Accordingly, there are now four disconnected components of the group O$(r,d-r, \mathbbm{R})$. For $d$ odd one can write again O$(r,d-r, \mathbbm{R})=Z_2 \times \text{SO}(r,d-r,\mathbbm{R})$ where the latter has only two disconnected components. Depending on whether the number of time dimensions $r$ or the number of space dimensions $(d-r)$ is odd, these two topologically disconnected components are connected by time reflections or space reflections, respectively. For $d$ even, the group is not of a simple product structure but still has four disconnected components in the real case. We denote the component topologically connected to the unit transformation by $\text{SO}^\uparrow(r,d-r, \mathbbm{R})$. This subgroup is actually a Lie group. Table \ref{tab:TopologyLorentzGroup} illustrates the topological structure of $O(r,d-r, \mathbbm{R})$ and decomposes it into four sectors I, II, III and IV.

Note that two subsequent transformations out of a single sector always lead to I. The structure is the one of the finite group $Z_2 \times Z_2$. In other words, $\text{O}(r,d-r, \mathbbm{R})/ \text{SO}^\uparrow(r,d-r, \mathbbm{R}) \cong Z_2 \times Z_2$. 

\paragraph{Space and time reflections.} For $d-r$ even, space reversion (i.\ e.\ the reflection along all time-like coordinate axis) ${\cal P}$ does not connect different topologically disconnected elements of the group but for $d-r$ odd this is the case. Similarly, when $r$ is even, simple time reversal ${\cal T}$ does not connect different components\footnote{By a {\it simple} time reflection we mean a discrete transformation that reverses the direction of all time coordinates but does {\it not} include a complex conjugation of all complex numbers $i \to -i$ as it would be the case for Wigner's anti-unitary time reversal in quantum mechanics.}, but  for $r$ odd, they do. Combined transformations ${\cal P T} = - \mathbbm{1}$ connect different components for $d$ odd or for $d$ even with $r$ and $d-r$ odd. In table \ref{tab:PTtransformations} we show the action of ${\cal P}$ and ${\cal T}$ in different dimensions.
\begin{table}
\centering
\begin{tabular}{ c | c | c }
\hline\hline
 & $(d-r)$ even & $(d-r)$ odd \\ \hline
 $r$ even & 
  \begin{tabular}{@{}c@{}} $d$ even \\ 
  ${\cal P} \in \text{I}, \quad  {\cal T} \in \text{I}, \quad {\cal PT} \in \text{I}$  \end{tabular} 
  & 
  \begin{tabular}{@{}c@{}} $d$ odd \\ 
  ${\cal P} \in \text{II}, \quad  {\cal T} \in \text{I}, \quad {\cal PT} \in \text{II}$  \end{tabular}      \\ \hline
$r$ odd 
& 
  \begin{tabular}{@{}c@{}} $d$ odd \\ 
  ${\cal P} \in \text{I}, \quad  {\cal T} \in \text{III}, \quad {\cal PT} \in \text{III}$  \end{tabular} & 
  \begin{tabular}{@{}c@{}} $d$ even \\ 
  ${\cal P} \in \text{II}, \quad  {\cal T} \in \text{III}, \quad {\cal PT} \in \text{IV}$ \end{tabular} \\
\hline\hline
\end{tabular} 
\caption{Action of space reflections ${\cal P}$ and simple time reflections ${\cal T}$ for different dimensions and signatures of the generalized Lorentz group O$(r,d-r, \mathbbm{R})$.}
\label{tab:PTtransformations}
\end{table}

\paragraph{Cartan-Dieudonn\'e theorem.} By virtue of the Cartan-Dieudonn\'e theorem one can compose all finite orthogonal transformations or elements of O$(r,d-r, \mathbbm{R})$ by a number (actually smaller or equal to $d$) of reflections along certain directions. We do not prove this interesting statement here, but just state it for later use. A proof can be found e.\ g.\ in refs.\ \cite{Cartan, ReeseHarvey}

\paragraph{Lie group.} As already mentioned, the connected subgroup $\text{SO}^\uparrow(r,d-r, \mathbbm{R})$ is a Lie group and may be discussed in terms of the Lie algebra. Infinitesimal transformations are of the form
\begin{equation}
\Lambda^\mu_{\;\;\nu} = \delta^\mu_{\;\;\nu} + \delta \omega^\mu_{\;\;\nu}.
\label{eq:LorentzTransformInfinitesimal}
\end{equation}
For $\delta\omega_{\mu\nu} = \eta_{\mu\rho}\delta\omega^\rho_{\;\;\nu}$ the condition \eqref{eq:GeneralizedLorentzGroup} implies anti-symmetry
\begin{equation}
\delta\omega_{\mu\nu} = - \delta\omega_{\nu\mu}.
\end{equation}
Representations of the Lorentz group with can be written in infinitesimal form as
\begin{equation}
U(\Lambda) = \mathbbm{1} + \frac{i}{2} \delta \omega_{\mu\nu} M^{\mu\nu},
\label{eq:defGeneratorM}
\end{equation}
and finite transformations as corresponding exponentiation. The generators are anti-symmetric, $M^{\mu\nu} = - M^{\nu\mu}$, and their Lie bracket is
\begin{equation}
\left[ M^{\mu\nu}, M^{\rho\sigma} \right] = i \left( \eta^{\mu\rho} M^{\nu\sigma} - \eta^{\mu\sigma} M^{\nu\rho} - \eta^{\nu\rho} M^{\mu\sigma} + \eta^{\nu\sigma} M^{\mu\rho} \right).
\label{eq:LorentzAlgebra}
\end{equation}
The fundamental representation \eqref{eq:LorentzTransformInfinitesimal} has the generators
\begin{equation}
(M_F^{\mu\nu})^\alpha_{\;\;\beta} = - i (\eta^{\mu\alpha}\delta^{\nu}_\beta - \eta^{\nu\alpha} \delta^\mu_\beta).
\label{eq:fundamentalGeneratorLorentz}
\end{equation}

\section{Real Clifford algebras}
 
Let us now introduce a real Clifford algebra $\mathcal{C}l(r,d-r, \mathbbm{R})$ by the following characterization. We take $\gamma^\mu$ to be the generators of an {\it associative algebra}\footnote{
An associative algebra is an algebraic structure with compatible operations of addition, multiplication (assumed to be associative), and a scalar multiplication by elements in some field, here $\mathbbm{R}$.} over $\mathbbm{R}$ satisfying
\begin{equation}
\gamma^\mu \gamma^\nu + \gamma^\nu \gamma^\mu = 2 \eta^{\mu\nu} \mathbbm{1},
\label{eq:defClifford}
\end{equation}
where $\mathbbm{1}$ is the unit element in the algebra and $\eta^{\mu\nu}=\text{diag}(-1,\ldots, -1, +1, \ldots, +1)$ is the (inverse) metric as introduced above. The algebra $\mathcal{C}l(r,d-r, \mathbbm{R})$ is then formed by arbitrary products of the generators $\gamma^\mu$ and {\it real} linear combinations thereof.

Note that the generators $\gamma^\mu$ are introduced here independent of a specific representation. We will later on construct specific representations in terms of matrices. As a side remark let us state here that the operators  $\theta^\mu+\eta^{\mu\alpha} \frac{\partial}{\partial \theta^\alpha}$ acting on a Grassmann algebra would form one such representation.

\paragraph{Sorted products and vector space.}
The elements of ${\cal C}l(r,d-r, \mathbbm{R})$ form a real vector space spanned by the unit element $\mathbbm{1}$ and the sorted products 
\begin{equation}
\gamma^{\mu_1 \mu_2 \cdots \mu_p} = \gamma^{\mu_1} \gamma^{\mu_2} \ldots \gamma^{\mu_p},
\end{equation}
with $\mu_1<\mu_2< \ldots < \mu_p$. More generally, we take the symbol $\gamma^{\mu_1 \mu_2 \cdots \mu_p}$ to be anti-symmetric under the permutation of any pair of neighboring indices and to vanish accordingly when two indices are equal.

\paragraph{Dimension of Clifford algebra.} Each generator can be present or absent in a sorted product, so the dimension of the Clifford algebra as a real vector space is $2^d$. (It could in principle be smaller if not all these elements are linearly independent. This case, which can specifically arise for $d$ odd and $r-(d-r) = 3 \text{ mod }4$ is usually excluded as a ``non universal'' Clifford algebra. Universal Clifford algebras have then dimension $2^d$.) 

\paragraph{Covectors.} The Clifford algebra contains in particular a copy of the dual vector space, or space of one-forms, with elements of the form 
\begin{equation}
v = v_\mu \gamma^\mu.
\end{equation}

\paragraph{$2$-Covectors and $p$-covectors.} One also has a copy of the space of two-forms, or $2$-covectors, with elements of the form
\begin{equation}
w = \sum_{\mu\nu}\frac{1}{2} w_{\mu\nu} \left(\gamma^\mu \gamma^\nu - \gamma^\nu \gamma^\mu \right) = \sum_{\mu < \nu} w_{\mu\nu} \gamma^{\mu\nu}.
\end{equation}
Note that one has here $w_{\mu\nu}=-w_{\nu\mu}$. In a similar way, also $p$-forms or $p$-covectors
\begin{equation}
u = \sum_{\mu_1 < \cdots < \mu_p} u_{\mu_1 \cdots \mu_p} \gamma^{\mu_1 \cdots \mu_p},
\end{equation}
with $u_{\mu_1 \cdots \mu_p}$ fully anti-symmetric, are embedded in the algebra. The entire exterior algebra is incorporated in this way in terms of combinations of generators in different directions. In fact, there is an isomorphism with the identification
\begin{equation}
\gamma^{\mu_1 \cdots \mu_p} \cong \text{d}x^{\mu_1} \wedge \text{d}x^{\mu_2} \wedge \ldots \wedge \text{d}x^{\mu_p}.
\end{equation}
However, note that $\gamma^\mu \gamma^\mu = \eta^{\mu\mu}$ while $dx^\mu \wedge dx^\mu = 0$. We will use this notation to write
\begin{equation}
\frac{1}{2} \left(\gamma^\mu \gamma^\nu - \gamma^\nu \gamma^\mu \right) = \gamma^\mu \wedge \gamma^\nu = \gamma^{\mu\nu}.
\end{equation}
Let us recall that $\gamma^{\mu\nu}$ is anti-symmetric under the permutation of indices.

\paragraph{Modified summation convention.}
In expressions involving $\gamma^{\mu_1 \cdots \mu_p}$ or the wedge product like $\gamma^\mu \wedge \gamma^\nu$ we use a modified summation convention where a restriction is imposed as above, such that e.\ g.\
\begin{equation}
w_{\mu\nu} \gamma^{\mu\nu} = \sum_{\mu<\nu} w_{\mu\nu} \gamma^{\mu\nu} = w_{\mu\nu} \gamma^{\mu} \wedge \gamma^\nu = \sum_{\mu<\nu} w_{\mu\nu} \gamma^{\mu} \wedge \gamma^\nu,
\end{equation}
and
\begin{equation}
u_{\mu_1 \cdots \mu_p} \gamma^{\mu_1 \cdots \mu_p}= \sum_{\mu_1 < \cdots < \mu_p} u_{\mu_1 \cdots \mu_p} \gamma^{\mu_1 \cdots \mu_p}.
\end{equation}

\paragraph{Decomposition into $p$-covectors.} We use the notation $\langle v \rangle_p$ for the $p$-covector part of a Clifford algebra element $v$. An arbitrary element of the algebra can be written as
\begin{equation}
\begin{split}
v = & v_{(0)} \mathbbm{1} + v_\mu \gamma^\mu + v_{\mu_1 \mu_2} \gamma^{\mu_1 \mu_2} + \ldots + v_{1-r\cdots d-r} \gamma^{1-r\cdots d-r} \\
= & \langle v \rangle_0 + \langle v \rangle_1 + \langle v \rangle_2 + \ldots + \langle v \rangle_d.
\end{split}
\label{eq:decompositionCliffordElement}
\end{equation}
Indeed, this corresponds to $1+d+d(d-1)/2 + \ldots + d+1 = 2^d$ linearly independent terms (for universal Clifford algebras).

\paragraph{Clifford product.} The Clifford product of two (covector) elements $u=u_\mu \gamma^\mu$ and $v=v_\nu \gamma^\nu$ is given by
\begin{equation}
u v = \frac{1}{2} \left( uv - vu \right) + \frac{1}{2} \left( uv + vu \right) = u_\mu v_\nu \gamma^{\mu\nu} + u_\mu v_\nu \eta^{\mu\nu} = u \wedge v + u \cdot v.
\end{equation}
This can be easily generalized to more general elements of the algebra, see e.\ g.\ refs.\ \cite{Lounesto:2001zz, Vaz:2016qyw}.

\paragraph{Clifford parity, grade involution and even subalgebra.} One can define an internal parity $\mathsf{G}$ in the Clifford algebra $\mathcal{C}l(r,d-r, \mathbbm{R})$ such that $\mathsf{G}(\gamma^\mu)=-\gamma^\mu$ and $\mathsf{G}(\langle v\rangle_p)=(-1)^p \langle v \rangle_p$. It splits the vector space $\mathcal{C}l(r,d-r, \mathbbm{R})$ into two vector spaces, $\mathcal{C}l_+(r,d-r, \mathbbm{R})$ and $\mathcal{C}l_-(r,d-r, \mathbbm{R})$, containing even and odd elements, respectively. One of them, namely $\mathcal{C}l_+(r,d-r, \mathbbm{R})$ containing only even elements is in fact a {\it subalgebra}. 

Acting on an element of the algebra, Clifford parity is also known as {\it grade involution} and one has
\begin{equation}
\begin{split}
\mathsf{G}(v) = \langle v \rangle_0 - \langle v \rangle_1 + \langle v \rangle_2 - \ldots + (-1)^d \langle v \rangle_d.
\end{split}
\end{equation}
Grade involution is an {\it automorphism} is the sense that
\begin{equation}
\mathsf{G}(uv) = \mathsf{G}(u) \mathsf{G}(v).
\end{equation}

\paragraph{Product of all generators.} One element of ${\cal C}l(r,d-r,\mathbbm{R})$ plays a distinguished role, the product of all generators (sometimes called ``volume element'')
\begin{equation}
\hat \gamma = \gamma^{(1-r)(2-r) \ldots 0 \ldots (d-r)} = \gamma^{1-r} \gamma^{2-r} \cdots \gamma^0 \gamma^1 \cdots \gamma^{d-r}.
\label{eq:defhatgamma}
\end{equation}
One can also write this as an expression where the order of generators is reversed,
\begin{equation}
\hat \gamma = (-1)^{d(d-1)/2} 
\gamma^{d-r} \cdots \gamma^1 \gamma^0 \cdots \gamma^{1-r}.
\label{eq:defhatgammaReversed}
\end{equation}
Note that to go from equation \eqref{eq:defhatgamma} to \eqref{eq:defhatgammaReversed} one needs to permute the first gamma matrix $d-1$ times and so on, and the last two matrices once. This makes
\begin{equation}
(d-1) + (d-2)+ \ldots + 1=d(d-1)/2
\end{equation}
permutations and corresponding signs. Moreover, one can write
$$
{d(d-1)/2} \text{ mod }2={(d+4)(d+4-1)/2} \text{ mod } 2 = [d/2] \text{ mod } 2,
$$
where $[d/2]$ is the integer part of $d/2$. 

One finds thus
\begin{equation}
\hat \gamma^2 = (-1)^{[d/2]-r} \mathbbm{1} = (-1)^{[((d-r)-r)/2]} \mathbbm{1} .
\label{eq:squaregammahat}
\end{equation}
Note that this depends only on the difference between the number of space and time directions $(d-r)-r \text{ mod }4$. Specifically, one has $\hat \gamma^2=\mathbbm{1}$ for $r-(d-r)=0,3 \text{ mod }4$ and $\hat \gamma^2= - \mathbbm{1}$ for $r-(d-r)=1,2 \text{ mod }4$.

The $d$-dimensional analogue of $\gamma_5$ can be defined by
\begin{equation}
\bar \gamma = (-i)^{[((d-r)-r)/2]} \hat\gamma,
\label{eq:hatgammadef}
\end{equation}
such that $\bar \gamma^2=1$. For $d=4$ and $r=1$ this agrees with the standard definition. Note that $\hat \gamma$ is an element of the {\it real} Clifford algebra over $\mathbbm{R}$ while $\bar\gamma$ is in general not. 

If $d$ is even, $\hat \gamma$ anti-commutes with the generators $\gamma^\mu$ and therefore with the elements of ${\cal C}l_-(r,d-r, \mathbbm{R})$ while it commutes with the elements of ${\cal C}l_+(r,d-r, \mathbbm{R})$. Instead, if $d$ is odd, $\hat \gamma$ commutes with all the generators and therefore {\it all} elements of the Clifford algebra ${\cal C}l(r,d-r,  \mathbbm{R})$.

\paragraph{Center of the algebra.} The center of the algebra, i.\ e.\ the set of elements which commutes with all other elements of ${\cal C}l(r,d-r,\mathbbm{R})$ contains only real multiples of the unit element $\mathbbm{1}$ for $d$ even. Instead, if $d$ is odd, the center is spanned by $\mathbbm{1}$ and $\hat\gamma$. 

\paragraph{Reversion.} In addition to grade involution $\mathsf{G}$ there is another natural involution $\mathsf{R}$ where one reverses the order of all generators, for example $\mathsf{R}(\gamma^\mu\gamma^\nu) = \gamma^\nu \gamma^\mu = - \gamma^\mu \gamma^\nu$ for $\mu\neq \nu$ etc. For a $p$-covector one has $\mathsf{R}(\langle v \rangle_p) = (-1)^{p(p-1)/2} \langle v \rangle_p$ and for an arbitrary element of the Clifford algebra
\begin{equation}
\begin{split}
\mathsf{R}(v) = \langle v \rangle_0 + \langle v \rangle_1 - \langle v \rangle_2 + \ldots + (-1)^{d(d-1)/2} \langle v \rangle_d.
\end{split}
\end{equation}
Reversion is an {\it anti-automorphism} in the sense that
\begin{equation}
\mathsf{R}(u v) = \mathsf{R}(v) \mathsf{R}(u).
\end{equation}
This is an immediate consequence of changing the order of generators.

\paragraph{Clifford conjugation.} Clifford conjugation $\mathsf{C}$ is the combination of grade involution $\mathsf{G}$ and reversion $\mathsf{R}$, 
\begin{equation}
\mathsf{C}(v) = \mathsf{R}(\mathsf{G}(v)) = \mathsf{G}(\mathsf{R}(v)).
\end{equation}
Specifically, for some element of the Clifford algebra,
\begin{equation}
\mathsf{C}(v) = \langle v \rangle_0 - \langle v \rangle_1 - \langle v \rangle_2 + \ldots + (-1)^{d(d+1)/2} \langle v \rangle_d.
\end{equation}
Again this is an anti-automorphism such that
\begin{equation}
\mathsf{C}(u v) = \mathsf{C}(v) \mathsf{C}(u).
\end{equation}

Grade involution, reversion and Clifford conjugation will be useful to define different involutions not only in the Clifford algebra and its representations, but also for associated spinors.

\paragraph{The structure map in the Clifford algebra.} One may ask whether grade involution can be realized as an $\mathbbm{R}$-linear map $\varsigma$ on the Clifford algebra such that for $a\in \mathcal{C}l(r,d-r, \mathbbm{R})$ one has
\begin{equation}
\mathsf{G}(a) = \varsigma \, a \, \varsigma^{-1}.
\label{eq:StructureMapCliffordAlgebra}
\end{equation}
Specifically, this needs $\varsigma \gamma^\mu \varsigma^{-1} = - \gamma^\mu$. Note that $\mathsf{G}(\mathsf{G}(a))=a$ and one can rescale $\varsigma$ such that $\varsigma^2=\pm \mathbbm{1}$. One can not demand $\varsigma^2=\mathbbm{1}$ in all cases, see below. In any case, this structure map $\varsigma$ is unique only up to a sign within a real Clifford algebra.

For $d$ even, $\hat \gamma$ as defined in \eqref{eq:defhatgamma} anti-commutes with all generators $\gamma^\mu$ and one can set $\varsigma = \hat \gamma$ or $\varsigma = -\hat \gamma$. The structure map $\varsigma$ is here actually itself part of the Clifford algebra, $\varsigma\in \mathcal{C}l(r,d-r,\mathbbm{R})$.  Note that $\varsigma^2=\hat \gamma^2 = \mathbbm{1}$ for $r-(d-r)=0,4 \text{ mod }8$ but $\varsigma^2=\hat \gamma^2 = - \mathbbm{1}$ for $r-(d-r) = 2,6$ mod $8$. 

For $d$ odd, $\hat \gamma$ commutes with all generators and is part of $\mathcal{C}l_-(r,d-r, \mathbbm{R})$. In fact, every element of $\mathcal{C}l_-(r,d-r, \mathbbm{R})$ can be written in the form of a product of an even element with $\hat \gamma$ and one has $\mathcal{C}l(r,d-r, \mathbbm{R}) = \mathcal{C}l_+(r,d-r, \mathbbm{R}) \oplus \mathcal{C}l_+(r,d-r, \mathbbm{R}) \hat \gamma$. The structure map $\varsigma$ corresponds to the map $\hat \gamma \to - \hat \gamma$. In this case $\varsigma$ is {\it not} itself an element of the Clifford algebra.

Let us now distinguish two cases. In the first case one has $r-(d-r)=1,5 \text{ mod }8$ and $\hat\gamma^2=-\mathbbm{1}$. Here $\hat\gamma$ defines a complex structure, typically $\hat\gamma=\pm i \mathbbm{1}$. One must take the structure map $\varsigma$ to be complex conjugation with respect to this complex structure. In addition, one may add an overall sign.

In the other case $r-(d-r)=3,7 \text{ mod }8$ and $\hat \gamma^2=\mathbbm{1}$. The Clifford algebra is now reducible. Usually $\hat \gamma$ acts on a direct sum of two algebras $A \oplus B$ as $(\mathbbm{1}, -\mathbbm{1})$.  The structure map interchanges the elements of the two summands, $a \oplus b \to b \oplus a$. Alternatively, $a\otimes b \to - b \otimes a$ works, as well.

In summary, depending on the dimension $d$ and the signature $r$, the grade involution $\mathsf{G}$ can be realized as a linear map which may (up to a sign) either be given by the product of all generators $\hat \gamma$ (for $d$ even), complex conjugation, or the interchange of the two summands when the Clifford algebra has the structure of a direct sum (for $d$ odd). 

\section{Pin and spin groups}
\label{sec:PinAndSpinGroups}

There is a very close relation between the orthogonal group O$(r,d-r, \mathbbm{R})$ and the so called pin group Pin$(r,d-r, \mathbbm{R})$ as well as between the special orthogonal group SO$(r,d-r, \mathbbm{R})$ and the spin group Spin$(r,d-r, \mathbbm{R})$ on the other side. We will now discuss this correspondence. Note that Clifford algebra elements are in general not invertible. To define groups, one must concentrate on those elements $a$ for which $a^{-1}$ exists.

\paragraph{Reflections along a vector.} We start by considering reflections along a specific covector direction $u_\mu$. One may define a projector orthogonal to this direction as $\delta_\mu^{\;\;\nu} - \frac{u_\mu u^\nu}{u \cdot u}$ and the projector in the direction of $u_\mu$ is obviously $\frac{u_\mu u^\nu}{u \cdot u}$. A transformation that reflects along the direction $u_\mu$ is accordingly given by
\begin{equation}
R_\mu^{\;\;\nu} = \delta_\mu^{\;\;\nu} - 2\frac{u_\mu u^\nu}{u \cdot u}.
\end{equation}
When acting on some vector $v_\nu$ and after contracting with Clifford algebra generators one finds
\begin{equation}
Rv=\gamma^\mu R_\mu^{\;\;\nu} v_\nu = v - \frac{2 \, u\cdot v}{u\cdot u} u = v - \frac{v u + u v}{u \cdot u} u = - u  v u^{-1} = \mathsf{G}(u) v u^{-1},
\end{equation}
where $u$ and $v$ are covector elements of the Clifford algebra. We also used $u^{-1} = u / (u \cdot u)$ and the grade involution $\mathsf{G}(u)=-u$ for a covector. We have thus constructed a representation of reflections along some axis in the Clifford algebra. By virtue of the Cartan-Dieudonn\'e theorem one can actually compose all orthogonal transformations by a number of reflections, so this is all we need.

Note that for a pure covector $v = v_\mu \gamma^\mu$ one has $\mathsf{R}(v) v = -\mathsf{C}(v) v = v \cdot v \in \mathbbm{R}$. Also for products of several pure covectors $v = v_1 \cdots v_N$ one has $\mathsf{R}(v) v = \pm \mathsf{C}(v)  v \in \mathbbm{R}$. Reflections along space-like directions can moreover be normalized to $\mathsf{R}(v) v = -\mathsf{C}(v) v = v \cdot v = +1$ and reflections along time-like directions to $\mathsf{R}(v) v = -\mathsf{C}(v) v = v \cdot v = -1$. Products of several normalized reflections $v = v_1 \cdots v_N$ satisfy then $\mathsf{R}(v) v = \pm 1$ and $\mathsf{C}(v) v = \pm 1$.

\paragraph{Clifford-Lipschitz group.} We first consider the so-called Clifford-Lipschitz group $\Gamma(r,d-r, \mathbbm{R})$ defined by
\begin{equation}
\Gamma(r,d-r, \mathbbm{R}) = \left\{ a \in \mathcal{C}l(r,d-r, \mathbbm{R}) \; | \;  \mathsf{G}(a) v a^{-1} \in V^*, \text{ for all } v=v_\mu \gamma^\mu \in V^* \right\}
\end{equation}
We use here the grade involution $\mathsf{G}(a)$ and the above corresponds to the so-called twisted adjoint representation. Because this is a linear transformation one must be able to write $\mathsf{G}(a) \gamma^\mu a^{-1} = M^\mu_{\;\; \nu} \gamma^\nu$ with some matrices $M^\mu_{\;\;\nu}$. Moreover, taking the grade involution of this expression gives $a \gamma^\mu \mathsf{G}(a)^{-1}=M^\mu_{\;\;\nu} \gamma^\nu$. We show now that $M^\mu_{\;\;\nu}$ is orthogonal,
\begin{equation}
M^\mu_{\;\;\nu} M^\rho_{\;\;\sigma} 2 \eta^{\nu\sigma} = M^\mu_{\;\;\nu} M^\rho_{\;\;\sigma} \left( \gamma^\nu \gamma^\sigma + \gamma^\sigma \gamma^\nu \right) = a \gamma^\mu \mathsf{G}(a^{-1} a) \gamma^\rho a^{-1} + a \gamma^\rho \mathsf{G}(a^{-1} a) \gamma^\mu a^{-1} =2 \eta^{\mu\rho}.
\end{equation}
As a consequence of the Cartan-Dieudonn\'e theorem, all orthogonal transformations are actually in $\Gamma(r,d-r, \mathbbm{R})$. This shows that there is a very close connection to the orthogonal group. Note that the correspondence in the above form works for any element of O$(r,d-r,\mathbbm{R})$. 

Moreover, if one restricts to even elements, 
\begin{equation}
\Gamma_+(r,d-r, \mathbbm{R}) = \Gamma(r,d-r, \mathbbm{R}) \cap \mathcal{C}l_+(r,d-r, \mathbbm{R}), 
\end{equation}
one obtains a subgroup that contains all elements of the special orthogonal group SO$(r,d-r, \mathbbm{R})$. To define this restricted group, one does not need the twisted adjoint representation as above but could equivalently use the adjoint representation $a v a^{-1}$.

Note, however, that the Clifford-Lipschitz group is in a certain sense larger than the orthogonal group, because group elements $a$ and $a^\prime = \lambda a$ that differ by a factor $0\neq \lambda\in \mathbbm{R}$ correspond to the same transformation $M^\mu_{\;\;\nu}$. This can be remedied by working with a normalized set of group elements.

\paragraph{Pin group.} The so-called pin group is defined as
\begin{equation}
\text{Pin}(r, d-r) = \left\{ a \in \Gamma(r,d-r, \mathbbm{R}) \; | \;  \mathsf{R}(a) a = \pm 1, \mathsf{C}(a) a = \pm 1  \right\}.
\label{eq:defPinGroup}
\end{equation}
It contains the elements of the Clifford-Lipschitz group, but they are now normalized.  Note that single reflections along a vector, and therefore products of such reflections, can be normalized in this way. By virtue of the Cartan-Dieudonn\'e theorem this covers all orthogonal transformations or elements of O$(r,d-r,\mathbbm{R})$. The degeneracy is now finite in the sense that only two elements $+a$ and $-a$ correspond to the same orthogonal transformation. One says that the pin group Pin$(r,d-r, \mathbbm{R})$ is the double coverage of the orthogonal group O$(r,d-r, \mathbbm{R})$.

Note that all elements of the pin group have a definite Clifford parity, $\mathsf{G}(a)=\pm a$. It is convenient to introduce the Clifford grade parity $\mathsf{g}(a)$ through
\begin{equation}
\mathsf{g}(a) = \begin{cases} 0 & \text{for } \mathsf{G}(a) = +a, \\ 1 & \text{for } \mathsf{G}(a) = -a. \end{cases}
\end{equation}
One can then write $\mathsf{G}(a) = (-1)^{\mathsf{g}(a)} a$ for any $a\in \text{Pin}(r,d-r, \mathbbm{R})$. With this notation, the action of an element of the pin group on a covector $v\in V^*$ can now be written alternatively as
\begin{equation}
v \to \mathsf{G}(a) v a^{-1} = a (-1)^{\mathsf{g}(a)} v a^{-1}.  
\end{equation}
Interestingly, this can now be easily generalized to an arbitrary product of generators. A product of an odd number of generators receives a sign, while a product of an even number of generators does not. For an arbitrary element of the Clifford algebra $b \in \mathcal{C}l(r,d-r,\mathbbm{R})$ one can write the transformation behavior under the pin group as
\begin{equation}
b \to a \, \mathsf{G}^{\mathsf{g}(a)}(b) \, a^{-1} = a \, \varsigma^{\mathsf{g}(a)} \, b \, (\varsigma^{-1})^{\mathsf{g}(a)} \, a^{-1}.
\label{eq:actionPinGroupCliffordAlgebraElement}
\end{equation}
We use here $\mathsf{G}^n=\mathsf{G}$ for $n$ odd and $\mathsf{G}^n=\mathbbm{1}$ for $n$ even. In the last equation we have used the structure map \eqref{eq:StructureMapCliffordAlgebra}. Eq.\ \eqref{eq:actionPinGroupCliffordAlgebraElement} will be particularly useful also in the context of spinors.

\paragraph{Restricted pin groups.} One can define a subgroups of the pin group by
\begin{equation}
\text{Pin}^\uparrow(r,d-r, \mathbbm{R}) = \left\{ a \in \Gamma(r,d-r, \mathbbm{R}) \; | \;\mathsf{R}(a) a = + 1  \right\}.
\end{equation}
This corresponds to the subgroup of the orthogonal transformations that do not change the orientation among the temporal coordinates because single reflections along time-like directions are excluded. Similarly, 
\begin{equation}
\text{Pin}^+(r,d-r, \mathbbm{R}) = \left\{ a \in \Gamma(r,d-r, \mathbbm{R}) \; | \;  \mathsf{C}(a) a = + 1  \right\},
\end{equation}
corresponds to the subgroup that does not change orientation among the spatial coordinates \cite{Vaz:2016qyw}. The action of these restricted pin groups on arbitrary elements of the Clifford algebra is of the form \eqref{eq:actionPinGroupCliffordAlgebraElement}. 

\paragraph{Spin group.} The spin group is obtained by restricting to even elements of the Clifford algebra, 
\begin{equation}
\text{Spin}(r, d-r) = \left\{ a \in \Gamma_+(r,d-r, \mathbbm{R}) \; | \; \mathsf{R}(a) a  =  \mathsf{C}(a) a  = \pm 1  \right\}.
\end{equation}
This group is the double coverage of SO$(r,d-r, \mathbbm{R})$. One can further restrict to  
\begin{equation}
\text{Spin}^\uparrow(r, d-r) = \left\{ a \in \Gamma_+(r,d-r, \mathbbm{R}) \; | \; \mathsf{R}(a) a = \mathsf{C}(a) a = + 1  \right\},
\label{eq:restrictedSpinGroup}
\end{equation}
which is now the double coverage of the Lie group SO$^\uparrow(r,d-r, \mathbbm{R})$.

Note that the elements of the spin group have obviously all even Clifford parity. In this sense the transformation law for an arbitrary element of the Clifford algebra is now also simpler than \eqref{eq:actionPinGroupCliffordAlgebraElement} and corresponds to the adjoint representation
\begin{equation}
b \to a \, b \, a^{-1}.
\label{eq:actionSpinGroupCliffordAlgebraElement}
\end{equation}

\paragraph{Lie group.} The subgroup $\text{Spin}^\uparrow(r, d-r)$ connected to the unit transformation is a  Lie groups and can be discussed in terms of its Lie algebra. Infinitesimal orthogonal transformations are specified in \eqref{eq:LorentzTransformInfinitesimal}, the corresponding commutation relations are given in \eqref{eq:LorentzAlgebra} and the fundamental representation is given in \eqref{eq:fundamentalGeneratorLorentz}. The commutation relations \eqref{eq:LorentzAlgebra} are also satisfied by the following generators in the Clifford algebra
\begin{equation}
M_S^{\mu\nu} = -\frac{i}{4} \left[ \gamma^\mu, \gamma^\nu \right].
\end{equation}
Exponentiation leads to a finite group element
\begin{equation}
L = \exp\left[ \frac{i}{2} \omega_{\mu\nu} M^{\mu\nu}_S \right].
\end{equation}
Note that rotations by $2\pi$ correspond to $L=-\mathbbm{1}$. This shows that the spin group and SO$^\uparrow(r,d-r, \mathbbm{R})$ are not isomorphic. One says that Spin$(r,d-r, \mathbbm{R})$ is the double cover of SO$^\uparrow(r,d-r, \mathbbm{R})$.
Note that $[\hat \gamma, M_S^{\mu\nu}] = 0$. For $d$ even where $\hat\gamma$ is non-trivial, one can therefore decompose the spin algebra generators $M_S^{\mu\nu}$ into left-handed and right-handed parts.

Interestingly, one can understand the Pin group \eqref{eq:defPinGroup} as consisting of transformations that are unitary or anti-unitary in the sense of $\mathsf{R}(a) a=\pm 1$ and $\mathsf{C}(a) a=\pm 1$.  This gets stepwise restricted to the restricted spin group \eqref{eq:restrictedSpinGroup} which contains only transformations that are unitary in both senses.

\paragraph{Non-orthogonal transformations.} One may ask whether also non-orthogonal transformations of spacetime can be realized as transformations in the Clifford algebra. Specifically, transformations of the type \eqref{eq:LorentzTransformInfinitesimal} correspond to orthogonal transformations for antisymmetric $\delta\omega_{\mu\nu}=-\delta\omega_{\nu\mu}$, while a symmetric part $\delta\omega_{\mu\nu}=\delta\omega_{\nu\mu}$ would describe a non-orthogonal transformation that also modifies the metric. One can convince oneself that such transformations can {\it not} be easily implemented in the Clifford algebra in the sense that one would find a transformation behavior for the generators which then gets generalized to other elements of the Clifford algebra by expressing them in terms of the generators. What one could do is to decompose the elements of the Clifford algebra into $p$-covectors as in eq.\ \eqref{eq:decompositionCliffordElement} and then to transform each element individually as a $p$-form.

In practice one circumvents this problem. Fermions are either described in flat cartesian coordinates (Minkowski space), or when this is not possible, for example in curved space one works with the tetrad (or vielbein) formalism and the spin connection, see e.\ g.\ refs.\ \cite{PenroseRindler, Lawson:1998yr}.

\section{Space and time reversal}

Now that we have found a direct correspondence between the orthogonal group O$(r,d-r, \mathbbm{R})$ and the Clifford group Pin$(r,d-r, \mathbbm{R})$, let us construct in particular space and time reversal transformations. The time reversal transformation we consider here is the so-called simple time reversal, in contrast to Wigner's time reversal which is realized as an anti-unitary transformation.

\paragraph{Time reversal $\mathcal{T}$.}
Simple time reversal is the reflection of all time-like coordinates and can be implemented by the product of all time-like generators
\begin{equation}
\beta = \gamma^{1-r} \cdots \gamma^0.
\label{eq:definitionBeta}
\end{equation}
Note that $\beta^2=\pm \mathbbm{1}$ depending on the number of time dimensions $r$. 
With these definitions one has the time reversal transformation in the twisted adjoint representation
\begin{equation}
\mathsf{G}(\beta) \gamma^\mu \beta^{-1} 
= \begin{cases} - \gamma^\mu & \text{for } (\gamma^\mu)^2 = -\mathbbm{1} \quad\quad (\mu \text{ time-like}) \\
+ \gamma^\mu & \text{for }  (\gamma^\mu)^2= + \mathbbm{1} \quad\quad (\mu \text{ space-like})
 \end{cases}
\end{equation}
Note that as transformation on the real Clifford algebra, $\beta$ is unique up to a sign. Note also that one has $\mathsf{G}(\beta) = \beta$ for $r$ even but $\mathsf{G}(\beta) = -\beta$ for $r$ odd.

For products of generators, the twisted adjoint presentation employed here has the consequence that elements of the odd and even sub algebra transform differently. More specific, for $a_+ \in \mathcal{C}l_+(r,d-r,\mathbbm{R})$ one has the time reversal 
\begin{equation}
\beta a_+ \beta^{-1},
\label{eq:timeReversal1}
\end{equation}
while for $a_-\in \mathcal{C}l_-(r,d-r,\mathbbm{R})$ one has the time reversal 
\begin{equation}
\mathsf{G}(\beta) a_- \beta^{-1}.
\label{eq:timeReversal2}
\end{equation}
For $r$ even this difference disappears because $\mathsf{G}(\beta) = \beta$ but for $r$ odd there is an additional minus sign in \eqref{eq:timeReversal2} compared to \eqref{eq:timeReversal1}. 

Using the representation of the pin group in eq.\ \eqref{eq:actionPinGroupCliffordAlgebraElement} one can summarize these transformation laws for time reversal for any $a\in \mathcal{C}l(r,d-r,\mathbbm{R})$ as
\begin{equation}
a \to \mathcal{T}(a) = \beta G^r(a) \beta^{-1} = \beta \varsigma^r a (\varsigma^{-1})^r \beta^{-1},
\label{eq:timeReversalTransformation}
\end{equation}
where $G^r=G$ for $r$ odd and $G^r=\mathbbm{1}$ for $r$ even. One may check that $\mathcal{T}(\mathcal{T}(a))=a$ such that $\mathcal{T}$ is an {\it involution}. Note, however, that $(\beta \varsigma^r)^2=\pm \mathbbm{1}$. 

\paragraph{Space reversal $\mathcal{P}$.}
Space reversal can be realized similarly by the product of all space-like generators
\begin{equation}
\alpha = \gamma^1 \cdots \gamma^{d-r}.
\label{eq:definitionBetaBar}
\end{equation}
Note that $\alpha^2=\pm \mathbbm{1}$, depending on the number of space dimensions $d-r$. 
Space reversal of generators is realized as
\begin{equation}
\mathsf{G}(\alpha) \gamma^\mu \alpha^{-1} 
= \begin{cases} + \gamma^\mu & \text{for } (\gamma^\mu)^2 = -\mathbbm{1} \quad\quad (\mu \text{ time-like}) \\
- \gamma^\mu & \text{for }  (\gamma^\mu)^2= + \mathbbm{1} \quad\quad (\mu \text{ space-like}).
 \end{cases}
\end{equation}
Again, as a transformation in the real Clifford algebra, $\alpha$ is unique up to a sign. One has $\mathsf{G}(\alpha)=\alpha$ for $(d-r)$ even while $\mathsf{G}(\alpha)=-\alpha$ for $(d-r)$ odd.

Again there is more generally a difference between even and odd elements of the Clifford algebra and one finds for $a\in \mathcal{C}l(r,d-r,\mathbbm{R})$ the space reversal transformation
\begin{equation}
a \to \mathcal{P}(a) = \alpha G^{d-r}(a) \alpha^{-1} = \alpha \varsigma^{d-r} a (\varsigma^{-1})^{d-r} \alpha^{-1},
\label{eq:SpaceReversalTransformation}
\end{equation} 
where $G^{d-r}=G$ for $d-r$ odd and $G^{d-r}=\mathbbm{1}$ for $d-r$ even. One may check that $\mathcal{P}(\mathcal{P}(a))=a$ such that $\mathcal{P}$ is an {\it involution} and $(\alpha \varsigma^{d-r})^2 = \pm \mathbbm{1}$. 

Let us note here that a combination of time reversal and space reversal corresponds to a reflection of all generators $\gamma^\mu \to -\gamma^\mu$. This is in fact precisely the grade involution $\mathsf{G}=\mathcal{T}\mathcal{P}$. One can also see this from the explicit realizations \eqref{eq:timeReversalTransformation} and \eqref{eq:TimeReversalSpinors} with the side information that $\beta\alpha=\hat \gamma$ and that $\varsigma=\pm \hat \gamma$ for $d$ even and that $\hat\gamma$ commutes with all generators when $d$ is odd. In short
\begin{equation}
\mathcal{T}(\mathcal{P}(a)) = \mathcal{P}(\mathcal{T}(a))  = \varsigma a \varsigma^{-1}.
\end{equation}

\section{More involutions}
We have already discussed three important involutions on the real Clifford algebra, namely grade involution $\mathsf{G}$, reversion $\mathsf{R}$ and Clifford conjugation $\mathsf{C}$. In addition to this, we have the discrete transformations of (simple) time reversal $\mathcal{T}$ and space reversal $\mathcal{P}$. In the present section we discuss further involution operations that are useful for physics applications in practice.

\subsection{Hermitian conjugation}
One may define another involution on the Clifford algebra. In the matrix representations we will consider below this corresponds to hermitian conjugate, so we use the notion $\mathsf{H}$ here.

Let us define the ``hermitian conjugate'' of a generator such that
\begin{equation}
\mathsf{H}(\gamma^\mu) = \begin{cases} + \gamma^\mu & \text{for } (\gamma^\mu)^2 =+\mathbbm{1}, \\
 - \gamma^\mu &\text{for } (\gamma^\mu)^2=-\mathbbm{1}.
\end{cases}
\label{eq:defHermitianConjugateGenerator}
\end{equation}
Note that generators in time-like directions are anti-hermitian in this sense, while those of space-like directions are hermitian. One may now extend this in a natural way to products of generators by requesting
\begin{equation}
\mathsf{H}(AB) = \mathsf{H}(B) \mathsf{H}(A).
\label{ABdagger}
\end{equation}
In this sense, $\mathsf{H}$ is an {\it anti-automorphism}. 

One has then for products of generators similar as for the generators themselves
\begin{equation}
\mathsf{H}(\gamma^{\mu_1\cdots \mu_p}) = \begin{cases} + \gamma^{\mu_1\cdots \mu_p} & \text{for } (\gamma^{\mu_1\cdots \mu_p})^2 =+\mathbbm{1}, \\
 - \gamma^{\mu_1\cdots \mu_p} &\text{for } (\gamma^{\mu_1\cdots \mu_p})^2=-\mathbbm{1}.
\end{cases}
\end{equation}
Moreover, the definition \eqref{eq:defHermitianConjugateGenerator} has the interesting consequence $\mathsf{H}(\gamma^\mu) \gamma^\mu=\mathbbm{1}$ and therefore the generators and any sequence of generators
\begin{equation}
A = \gamma^{\mu_1} \cdots \gamma^{\mu_n},
\end{equation}
are unitary in the sense $\mathsf{H}(A) A = \mathbbm{1}$.

\paragraph{Hermitian conjugate of structure map.} In the following we will also need the hermitian conjugate of the structure map $\varsigma$ defined in \eqref{eq:StructureMapCliffordAlgebra}. For $d$ even, this is easily found, because $\varsigma=\pm \hat \gamma$ is itself part of the Clifford algebra. Moreover, as a product of generators we have $\mathsf{H}(\hat \gamma) \hat \gamma = \mathbbm{1}$ and therefore 
\begin{equation}
\mathsf{H}(\varsigma) = \varsigma^{-1}.
\label{eq:HermitianConjugateCliffordStructure}
\end{equation}
For $d$ odd this is more involved because $\varsigma$ is not an element of the Clifford algebra. However, one can consistently extend the definitions such that \eqref{eq:HermitianConjugateCliffordStructure} holds also there. The consistency of this setting can be checked in concrete matrix representations of the odd-dimensional Clifford algebras.

\paragraph{Relation of hermitian conjugation $\mathsf{H}$ to other involutions.} Let us relate the involution $\mathsf{H}$ to the involutions discussed previously. In particular, for a single generator, it can be directly related to time reversal and space reversal,
\begin{equation}
\mathsf{H}(\gamma^\mu) = \mathsf{G}(\beta) \mathsf{R}(\gamma^\mu) \beta^{-1} =  \mathsf{G} (\alpha) \mathsf{C}(\gamma^\mu) \alpha^{-1}.
\label{eq:gammadaggerbeta}
\end{equation}
We have employed here the reverse $\mathsf{R}$ and Clifford conjugate $\mathsf{C}$ which are also anti-automorphisms. One can then extend this to arbitrary elements of the Clifford algebra $a\in \mathcal{C}l(r,d-r,\mathbbm{R})$ as
\begin{equation}
\begin{split}
\mathsf{H}(a) & = \beta \varsigma^{r-1} \mathsf{C}(a) (\varsigma^{-1})^{r-1} \beta^{-1} =  \beta \varsigma^r \mathsf{R}(a) (\varsigma^{-1})^r \beta^{-1} \\
& = \alpha \varsigma^{d-r} \mathsf{C}(a) (\varsigma^{-1})^{d-r} \alpha^{-1} 
= \alpha \varsigma^{d-r+1} \mathsf{R}(a) (\varsigma^{-1})^{d-r+1} \alpha^{-1},
\end{split}
\label{eq:HermitianConjugationGeneral}
\end{equation}
where we have used the structure map \eqref{eq:StructureMapCliffordAlgebra}.

\subsection{Dirac adjoints} 
We define now two more involutions on the Clifford algebra which are in fact rather useful for applications to relativistic fermions. They are not independent of the involutions introduced previously, though.

\paragraph{First Dirac adjoint.}
We define the first Dirac adjoint for $a\in \mathcal{C}l(r,d-r,\mathbbm{R})$ as the combination of hermitian conjugation and space reversal,
\begin{equation}
\mathsf{D}_1(a) = \mathcal{P}^{-1}(\mathsf{H}(a)) =  (\varsigma^{-1})^{d-r} \alpha^{-1} \, \mathsf{H}(a) \,\alpha \varsigma^{d-r} .
\label{eq:Dirac1GeneralClifford}
\end{equation}
Note that this is an {\it anti-automorphism} in the sense that $\mathsf{D}_1(ab) = \mathsf{D}_1(b) \mathsf{D}_1(a)$. Generators transform like $\mathsf{D}_1(\gamma^\mu) = - \gamma^\mu$. This, or inspection of \eqref{eq:HermitianConjugationGeneral} shows that the first Dirac adjoint agrees in fact with the Clifford conjugate,
\begin{equation}
\mathsf{D}_1(a) = \mathsf{C}(a).
\end{equation}
Note that in Minkowski space where $d=4$ and $r=1$ one has $\varsigma = \pm \hat \gamma=\pm \beta\alpha$ and
\begin{equation}
D_1(a) = \beta^{-1} \mathsf{H}(a) \beta = (\gamma^0)^{-1} \mathsf{H}(a) \gamma^0.
\end{equation}

\paragraph{Second Dirac adjoint.}
Similarly, a second Dirac adjoint for an element of the Clifford algebra $a\in \mathcal{C}l(r,d-r,\mathbbm{R})$ as the combination of hermitian conjugation and a space reversal,
\begin{equation}
\mathsf{D}_2(a) = \mathcal{T}^{-1}(\mathsf{H}(a)) = (\varsigma^{-1})^r \beta^{-1} \, \mathsf{H}(a) \,  \beta \varsigma^{r}.
\label{eq:Dirac2GeneralClifford}
\end{equation}
Again this is an {\it anti-automorphism}. Generators transform now as $\mathsf{D}_1(\gamma^\mu)=\gamma^\mu$ which shows that the second Dirac adjoint is in fact the reverse,
\begin{equation}
\mathsf{D}_1(a) = \mathsf{R}(a).
\end{equation}

\section{Matrix representations and classification}
\label{sec:MatrixRepresentations}

Here we will construct and discuss concrete representations for the Clifford algebra $\mathcal{C}l(r,d-r, \mathbbm{R})$ as a matrix algebra over $\mathbbm{R}$. This will naturally lead to real, complex and quaternionic structures that can be used to construct spinors for Majorana fermions and generalizations. 

The representation discussed in the present section is particularly useful to make real, complex and quaternionic structures explicit. In addition to discussing the Clifford algebras $\mathcal{C}l(r,d-r,\mathbbm{R})$, we will also discuss the possibilities for idempotents, which are elements $p \in \mathcal{C}l(r,d-r,\mathbbm{R})$  such that $p^2=p$. This will be beneficial for our later discussion of spinors as elements of (minimal) left ideals. To prepare this discussion we first review the concept of idempotents.

\subsection{Idempotents}
\label{sec:Idempotents}

\paragraph{Idempotent or projector.} A non-zero element of a Clifford algebra $p\in \mathcal{C}l$ is called an {\it idempotent} or a {\it projector} if $p^2=p$. Obviously this implies also $p^N=p$.

There is always the {\it trivial idempotent} $p=\mathbbm{1}$. If the algebra $\mathcal{C}l$ is isomorphic to a division algebra, such as $\mathbbm{R}$, $\mathbbm{C}$ or $\mathbbm{H}$, then the unique idempotent is the identity $\mathbbm{1}$. However, this is oftentimes not the case.

\paragraph{Orthogonal or annihlating idempotents.} Two projectors or idempotents are called {\it orthogonal} or {\it annihilating} if $p_1 p_2 = p_2 p_1 = 0$. 

\paragraph{Primitive or minimal idempotents.}
A projector or idempotent $p$ is said to be {\it primitive} or {\it minimal} if it cannot be written as the sum of two orthogonal idempotents $p\neq p_1+p_2$ where $p_1^2=p_1$, $p_2^2=p_2$ and $p_1 p_2 = p_2 p_1 = 0$. For a matrix algebra, the projector to a single coordinate $(p_i)_{jk} = \delta_{ji} \delta_{ki}$ would be an example for a primitive idempotent.

\paragraph{Quasi-minimal idempotent.} We will also introduce the notion of a {\it quasi-minimal idempotent}. It occurs for spaces that are in fact a direct sum denoted by $A\oplus B$ with elements $a\oplus b$ or simply $(a,b)$, where the structure map $\varsigma$ interchanges these two summands, $(a,b) \to (b,a)$ (up to a possible overall sign). A minimal idempotent $p_\text{m}$ is necessarily non-zero only in one of these two summands and therefore not mapped to itself by the structure map $\varsigma$. From such a minimal idempotent one can obtain a {\it quasi-minimal idempotent} $p_\text{q} = p_\text{m} + \varsigma p_\text{m}$ which is then by construction mapped to itself by the Clifford structure map $\varsigma$. It is arguably a minimal idempotent with this property.

\paragraph{Division ring from primitive idempotent.} For a primitive projector or  idempotent $p$, the subset of the Clifford algebra $p\, \mathcal{C}l \, p$ is in fact a division ring and it is isomorphic to $\mathbbm{R}$, $\mathbbm{C}$ or $\mathbbm{H}$, see e.\ g.\ ref.\ \cite{Vaz:2016qyw} for a proof.

\subsection{Real Clifford algebra representations up to two dimensions}

We now construct matrix representations for the real Clifford algebras $\mathcal{C}l(r,d-r,\mathbbm{R})$. We start from the low-dimensional cases which we discuss explicitly. This will also form a basis for an inductive construction of higher dimensional cases in terms of tensor products.

\paragraph{One dimension $d=1+0$.} The Clifford algebra $\mathcal{C}l(1,0, \mathbbm{R})$ has one single generator $\gamma^0$ with $(\gamma^{0})^2=-\mathbbm{1}$. The elements of the algebra are of the form $a \mathbbm{1}+b \gamma^0$ with $a,b \in \mathbbm{R}$. One may define a {\it complex structure} (a kind of complex conjugation) as $\gamma^0 \to - \gamma^0$. This shows that there is an isomorphism $\mathcal{C}l(1,0) \cong \mathbbm{C}$. In fact one could simply take $\gamma^0=i$. Grade involution $\mathsf{G}$ and Clifford conjugation $\mathsf{C}$ correspond to complex conjugation $\gamma^0 \to - \gamma^0$ or $i\to -i$, while reversion $\mathsf{R}$ has no effect here.

Because it must fulfill $p^2=p$, there is only the trivial idempotent in the algebra $\mathcal{C}l(1,0, \mathbbm{R})$, namely the unit element $p=\mathbbm{1}$.

\paragraph{One dimension $d=0+1$.} The Clifford algebra $\mathcal{C}l(0,1, \mathbbm{R})$ has one generator $\gamma^1$ with $(\gamma^1)^2=\mathbbm{1}$. The elements are of the form $a \mathbbm{1} + b \gamma^1$. One my represent the algebra by two-dimensional diagonal matrices 
\begin{equation}
a \mathbbm{1} + b \gamma^1 = \begin{pmatrix}  a+b & 0 \\ 0 & a-b\end{pmatrix}.
\end{equation}
This shows that there is an isomorphism to the algebra of diagonal two-by-two matrices, $\mathcal{C}l(0,1, \mathbbm{R}) \cong \mathbbm{R} \oplus \mathbbm{R}$. One may also write the elements as
\begin{equation}
a \mathbbm{1} + b \gamma^1 = a (1,  1) + b(1,-1)=(a+b, a-b).
\end{equation}
This representation is reducible. There are smaller, irreducible representations, or ``non-universal'' algebras, where $\hat \gamma=1$ or $\hat \gamma=-1$ and they are isomorphic to $\mathbbm{R}$. Grade involution $\mathsf{G}$ and Clifford conjugation $\mathsf{C}$ interchange the two copies of $\mathbbm{R}$ in the sense $(a+b,a-b) \to (a-b,a+b)$, while reversion $\mathsf{R}$ has no effect.

The algebra $\mathcal{C}l(0,1, \mathbbm{R})$ has the non-trivial idempotents $( \mathbbm{1}+ \gamma^1)/2$ and $(\mathbbm{1}- \gamma^1)/2$ corresponding to $(1,0)$ and $(0,1)$ respectively. These two are minimal idempotents, while the trivial idempotent $\mathbbm{1}=(1,1)$ is non-minimal (but in fact quasi-minimal according to the definition in section \ref{sec:Idempotents}).

We note in particular that the algebras $\mathcal{C}l(1,0, \mathbbm{R})$ and $\mathcal{C}l(0,1, \mathbbm{R})$ are not isomorphic. In both cases, the subalgebra of even elements $\mathcal{C}l_+$ is generated by $\mathbbm{1}$ only, and simply given by $\mathbbm{R}$. On the even subalgebra, grade involution $\mathsf{G}$, Clifford conjugation $\mathsf{C}$ and reversion $\mathsf{R}$ all have no effect.

\paragraph{Two dimensions $d=1+1$.} We now move to two dimensions and start with $\mathcal{C}l(1,1, \mathbbm{R})$.  We need two anti-commuting generators $\gamma^0$ and $\gamma^1$ with $-(\gamma^0)^2=(\gamma^1)^2=\mathbbm{1}$. A possible choice in terms of real matrices is
\begin{equation}
\gamma^0 = -i\sigma_2 = \begin{pmatrix} 0 & -1 \\ 1 & 0 \end{pmatrix}, \quad\quad\quad \gamma^1 = \sigma_1 = \begin{pmatrix} 0 & 1 \\ 1 & 0 \end{pmatrix}.
\end{equation}
Besides these and $\mathbbm{1}$, the forth linearly independent element of the algebra is given by
\begin{equation}
\hat \gamma = \gamma^0 \gamma^1 = - \sigma_3 = \begin{pmatrix} -1 & 0 \\ 0 & 1 \end{pmatrix}.
\end{equation}
Note that these matrices form a basis for the real vector space, and associative algebra, of real two-by-two matrices $\text{Mat}(2, \mathbbm{R})$. In other words, there is an isomorphism $\mathcal{C}l(1,1, \mathbbm{R}) \cong \text{Mat}(2, \mathbbm{R})$.  

This Clifford algebra has a rather large class of non-trivial idempotents. In fact, any $p=(\mathbbm{1} + a \gamma^0 + b\gamma^1 + c \hat \gamma)/2$ with $-a^2+b^2+c^2=1$ fulfills the requirements and is also minimal. Particularly simple choices are the {\it canonical idempotents} $(\mathbbm{1} \pm \hat \gamma)/2$.

The subalgebra of even elements $\mathcal{C}l_+(1,1, \mathbbm{R})$ is generated by the diagonal matrices $\mathbbm{1}$ and $\hat \gamma$ where $\hat\gamma^2=\mathbbm{1}$. One has therefore $\mathcal{C}l_+(1,1, \mathbbm{R}) \cong \mathcal{C}l(0,1,\mathbbm{R}) \cong \mathbbm{R} \oplus \mathbbm{R}$. 

\paragraph{Two dimensions $d=0+2$.} Now let us consider the Euclidean case $\mathcal{C}l(0,2, \mathbbm{R})$. We can take the generators
\begin{equation}
\gamma^1 = \sigma_1 = \begin{pmatrix} 0 & 1 \\ 1 & 0 \end{pmatrix}, \quad\quad\quad \gamma^2 = \sigma_3 = \begin{pmatrix} 1 & 0 \\ 0 & -1 \end{pmatrix}. 
\end{equation}
The forth linearly independent element is
\begin{equation}
\hat \gamma = \gamma^1 \gamma^2 = - i \sigma_2 = \begin{pmatrix} 0 & -1 \\ 1 & 0 \end{pmatrix}.
\end{equation}
Note that again these matrices form a basis for the real two-by-two matrices and thus there is an isomorphism $\mathcal{C}l(0,2, \mathbbm{R}) \cong \text{Mat}(2, \mathbbm{R})$. 

Again there is a rather large set of non-trivial idempotents. Any combination $p=(\mathbbm{1}+a\gamma1+b\gamma^2+c \hat \gamma)$ with $a^2+b^2-c^2=1$ fulfills $p^2=p$ and they are also minimal. Particularly simple are the canonical choices $(\mathbbm{1} \pm \gamma^2)/2$. 

The subalgebra of even elements $\mathcal{C}l_+(0,2, \mathbbm{R})$ is now generated by $\hat \gamma=-i \sigma_2$ which squares to $(\hat\gamma)^2=-\mathbbm{1}$. It constitutes a {\it complex structure} and one has the isomorphism between algebras $\mathcal{C}l_+(0,2, \mathbbm{R})\cong \mathcal{C}l(1,0,\mathbbm{R}) \cong \mathbbm{C}$. 

\paragraph{Two dimensions $d=2+0$.} Finally we consider $\mathcal{C}l(2,0, \mathbbm{R})$. We need now two generators that square to minus one, $(\gamma^{-1})^{2} = (\gamma^0)^2=-\mathbbm{1}$, so that they both constitute complex structures, and they should anti-commute. This is not possible within the real two-by-two matrices any more. A possible choice is 
\begin{equation}
\gamma^{-1} = -i \sigma_1 = \begin{pmatrix} 0 & -i \\ -i & 0 \end{pmatrix}, \quad\quad\quad \gamma^{0} = -i \sigma_2 = \begin{pmatrix} 0 & -1 \\ 1 & 0 \end{pmatrix}.
\label{eq:generators20a}
\end{equation}
The forth linearly independent element is
\begin{equation}
\hat \gamma = \gamma^{-1} \gamma^0 = -i \sigma_3 = \begin{pmatrix} - i & 0 \\ 0 & i \end{pmatrix}.
\label{eq:generators20b}
\end{equation}
Note that it also squares to minus one, $(\hat\gamma)^2=-\mathbbm{1}$. In this case there is now an isomorphism to the quaternion algebra $\mathbbm{H}$ with the identification $(1, \mathbf{i},\mathbf{j},\mathbf{k}) \cong (\mathbbm{1}, -i \sigma_1, -i\sigma_2, -i\sigma_3) \cong (\mathbbm{1}, \gamma^{-1}, \gamma^0, \hat \gamma)$. In other words, $\mathcal{C}l(2,0, \mathbbm{R}) \cong \mathbbm{H}$. Clifford conjugation $\mathsf{C}$ corresponds here to the quaternion conugate $(1,\mathbf{i},\mathbf{j}, \mathbf{k}) \to (1,-\mathbf{i},-\mathbf{j},-\mathbf{k})$ or hermitean conjugation $A\to A^\dagger$ in the above matrix realization. 

One may check directly that there is no non-trivial idempotent within the real Clifford algebra $\mathcal{C}l(2,0, \mathbbm{R})$. This follows also from the isometry with $\mathbbm{H}$ which is a division algebra. (Note that if complex coefficients were allowed, this statement would change and $(\mathbbm{1}\pm i \hat \gamma)/2$ would be examples for non-trivial and minimal idempotents.)

The even elements are generated by $\hat \gamma$ with $\hat\gamma^2=-\mathbbm{1}$, which is also a complex structure such that $\mathcal{C}l_+(2,0, \mathbbm{R}) \cong \mathcal{C}l(1,0,\mathbbm{R}) \cong \mathbbm{C}$.

\subsection{Inductive construction of real algebra representations for $d>2$}
\label{sec:InductiveConstruction}
On the basis of the Clifford algebras for $d\leq 2$ we construct now the higher dimensional cases in terms of an iterative tensor product construction.

\paragraph{Inductive construction.} From the one- and two-dimensional Clifford algebras one can construct all other cases by the following relations
\begin{equation}
\begin{split}
(i) \quad\quad & \mathcal{C}l(d,0, \mathbbm{R}) \otimes \mathcal{C}l(0,2, \mathbbm{R}) \cong \mathcal{C}l(0,d+2, \mathbbm{R}) \\
(ii) \quad\quad & \mathcal{C}l(0,d, \mathbbm{R}) \otimes \mathcal{C}l(2,0, \mathbbm{R}) \cong \mathcal{C}l(d+2, 0, \mathbbm{R}) \\
(iii) \quad\quad & \mathcal{C}l(r,d-r, \mathbbm{R}) \otimes \mathcal{C}l(1,1, \mathbbm{R}) \cong \mathcal{C}l(r+1, d-r+1, \mathbbm{R}) \\
\end{split} \label{eq:recursionCliffordAlgebras}
\end{equation}
To see this, assume that the generators  $\gamma^{\mu}_{(d,0)}$ of $\mathcal{C}l(d,0, \mathbbm{R})$ and $\gamma_{(0,2)}^\mu$ of $\mathcal{C}l(0,2, \mathbbm{R})$ are given. One can then set (recall that for the tensor product of algebras $(a_1 \otimes b_1) (a_2 \otimes b_2) = a_1 a_2 \otimes b_1 b_2$)
\begin{equation}
\begin{split}
\gamma^\mu_{(0,d+2)} = \begin{cases}
    \gamma^{\mu-d}_{(d,0)} \otimes \gamma_{(0,2)}^1 \gamma_{(0,2)}^2 & \text{for}\quad 1\leq \mu\leq d\\
    \mathbbm{1}_{(d,0)} \otimes \gamma_{(0,2)}^{\mu-d} & \text{for}\quad d+1\leq \mu \leq d+2.
  \end{cases}  
\end{split}
\end{equation}
In a similar way, if one takes $\gamma^{\mu}_{(0,d)}$ and $\gamma^{\mu}_{(2,0)}$ as given, one can set
\begin{equation}
\begin{split}
\gamma^\mu_{(d+2,0)} = \begin{cases}
    \gamma^{\mu+d+2}_{(0,d)} \otimes \gamma_{(2,0)}^{-1} \gamma_{(2,0)}^0 & \text{for}\quad -d-1\leq \mu\leq -2\\
    \mathbbm{1}_{(0,d)} \otimes \gamma_{(2,0)}^{\mu} & \text{for}\quad -1\leq \mu \leq 0.
  \end{cases}  
\end{split}
\end{equation}
Finally, if the generators $\gamma^\mu_{(r,d-r)}$ and $\gamma^\mu_{(1,1)}$ are given one can set
\begin{equation}
\begin{split}
\gamma^\mu_{(r+1,d-r+1)} = \begin{cases}
    \gamma^{\mu+1}_{(r,d-r)} \otimes \gamma_{(1,1)}^0 \gamma_{(1,1)}^1 & \text{for}\quad -r\leq \mu\leq -1\\
    \mathbbm{1}_{(r,d-r)} \otimes \gamma_{(1,1)}^{0} & \text{for}\quad \mu = 0 \\
    \gamma^{\mu}_{(r,d-r)} \otimes \gamma_{(1,1)}^0 \gamma_{(1,1)}^1 & \text{for}\quad 1\leq \mu\leq d-r\\
    \mathbbm{1}_{(r,d-r)} \otimes \gamma_{(1,1)}^{1} & \text{for}\quad \mu = d-r+1.    
  \end{cases}  
\end{split}
\end{equation}
Starting from $\mathcal{C}l(1,0, \mathbbm{R})$, $\mathcal{C}l(0,1, \mathbbm{R})$, $\mathcal{C}l(2,0, \mathbbm{R})$ and $\mathcal{C}l(0,2, \mathbbm{R})$ one can use relations $(i)$ and $(ii)$ in \eqref{eq:recursionCliffordAlgebras} to construct the purely timelike and purely spacelike Clifford algebras $\mathcal{C}l(d,0, \mathbbm{R})$ and $\mathcal{C}l(0,d, \mathbbm{R})$. Using then also relation $(iii)$ one can obtain the mixed cases. 

It is sometimes convenient to change the order of the tensor product in order to construct a concrete matrix representation. We will do so deliberately.

Note that the representation above is not at all unique. One could change the order of the gamma matrices and one can start from different representations of the low dimensional algebras. We will not use this construction for a concrete realization but for a qualitative classification of Clifford algebras.

In order to find (minimal) idempotents from the iterative construction \eqref{eq:recursionCliffordAlgebras} one can choose such idempotents in each of the tensor product factors. 

\paragraph{Classification of Clifford algebras.} From \eqref{eq:recursionCliffordAlgebras} one sees that a step of the type $(iii)$ which goes up in $r$ and $d-r$ simultaneously, and therefore does not change $r-(d-r)$, makes the dimension of the representation a factor $2$ larger but does not change the qualitative structure. 
\begin{table}
\centering
\begin{tabular}{ c | c | c }
\hline\hline
$r-(d-r) \text{ mod } 8$ & Isomorphic matrix algebra & Dimension  
\\ \hline
 0, 6 & $\text{Mat}(N, \mathbbm{R})$ & $N=2^{d/2}$  
 \\ \hline
 2, 4 & $\text{Mat}(N, \mathbbm{H})$ & $N=2^{(d-2)/2}$
 \\ \hline
 1, 5 & $\text{Mat}(N, \mathbbm{C})$ & $N=2^{(d-1)/2}$ 
 \\ \hline
 3 & $\text{Mat}(N/2, \mathbbm{H}) \oplus \text{Mat}(N/2, \mathbbm{H})$ & $N=2^{(d-1)/2}$ 
 \\ \hline
 7 & $\text{Mat}(N/2, \mathbbm{R}) \oplus \text{Mat}(N/2, \mathbbm{R})$ & $N=2^{(d+1)/2}$
 \\ \hline\hline
\end{tabular}
\caption{Classification of real Clifford algebras $\mathcal{C}l(r,d-r, \mathbbm{R})$ in terms of isomorphic real, complex or quaternionic matrix algebras.}
\label{tab:CliffordAlgebraClassification}
\end{table}

In particular if one takes $\mathcal{C}l(1,1, \mathbbm{R})\cong \text{Mat}(2,\mathbbm{R})$ as a starting point, one finds that the Clifford algebras $\mathcal{C}l(p,p, \mathbbm{R})$ are isomorphic to the real matrices of dimension $N=2^p$. Similarly, taking $\mathcal{C}l(1,0, \mathbbm{R})\cong \mathbbm{C}$ as a starting point shows that the Clifford algebras $\mathcal{C}l(1+p,p, \mathbbm{R})$ are isomorphic to $\text{Mat}(2^p, \mathbbm{C})$. In this way one can continue and one finds the following classification.
\begin{equation}
\begin{split}
\mathcal{C}l(p,p, \mathbbm{R}) & \cong \text{Mat}(2^p, \mathbbm{R}) \\
\mathcal{C}l(1+p,p, \mathbbm{R}) & \cong \text{Mat}(2^p, \mathbbm{C}) \\
\mathcal{C}l(p,1+p, \mathbbm{R}) & \cong \text{Mat}(2^p, \mathbbm{R}) \oplus \text{Mat}(2^p, \mathbbm{R}) \\
\mathcal{C}l(2+p,p, \mathbbm{R}) & \cong \text{Mat}(2^p, \mathbbm{H}) \\
\mathcal{C}l(p,2+p, \mathbbm{R}) & \cong \text{Mat}(2^{p+1}, \mathbbm{R}) \\
\mathcal{C}l(3+p,p, \mathbbm{R}) & \cong \text{Mat}(2^{p}, \mathbbm{H}) \oplus \text{Mat}(2^{p}, \mathbbm{H}) \\
\mathcal{C}l(p,3+p, \mathbbm{R}) & \cong \text{Mat}(2^{p+1}, \mathbbm{C}) \\
\mathcal{C}l(4+p,p, \mathbbm{R}) & \cong \text{Mat}(2^{p+1}, \mathbbm{H}).
\end{split}
\end{equation}
These relations can be extended using the mod $8$ periodicity in signature $r-(d-r)$ but keeping the dimension $d$ fixed, for example
\begin{equation}
\mathcal{C}l(p,4+p, \mathbbm{R}) \cong \mathcal{C}l(4+p,p, \mathbbm{R}) \cong \text{Mat}(2^{p+1}, \mathbbm{H})
\end{equation}
or
\begin{equation}
\mathcal{C}l(p,5+p, \mathbbm{R}) \cong \mathcal{C}l(3+1+p,1+p, \mathbbm{R}) \cong \text{Mat}(2^{p+1}, \mathbbm{H}) \oplus \text{Mat}(2^{p+1}, \mathbbm{H}).
\end{equation}
In this sense, the above classification is actually complete. Table \ref{tab:CliffordAlgebraClassification} summarizes the classification of real Clifford algebras. 

\paragraph{Even Clifford sub algebra.} It is also interesting to investigate the structure of the even subalgebra $\mathcal{C}l_+(r,d-r, \mathbbm{R})$. Let us first establish an isomorphism to the full Clifford algebra of lower dimension,
\begin{equation}
\begin{split}
\mathcal{C}l_+(r,d-r, \mathbbm{R}) \cong \mathcal{C}l(r-1,d-r, \mathbbm{R}) & \quad\quad\quad\text{for } r\geq1, \\
\mathcal{C}l_+(r,d-r, \mathbbm{R}) \cong \mathcal{C}l(d-r-1, s, \mathbbm{R}) & \quad\quad\quad\text{for } d-r\geq1.
\end{split}\label{eq:CLplusisomorphism}
\end{equation}
To see this, consider first $r>1$ and take $\gamma^{1-r}\cdots \gamma^{d-r}$ to be the generators of $\mathcal{C}l(r,d-r, \mathbbm{R})$. One may take $\gamma^{1-r}\gamma^\mu $ with $2-r\leq \mu \leq d-r$ as generators for the even subalgebra $\mathcal{C}l_+(r,d-r, \mathbbm{R})$. At the same time they are generators of $\mathcal{C}l(r-1,d-r, \mathbbm{R})$. In a similar way also the second line in \eqref{eq:CLplusisomorphism} can be understood. From table \ref{tab:CliffordAlgebraClassification} one obtains immediately a similar classification of even subalgebras in table \ref{tab:EvenCliffordAlgebraClassification}.

Note that it follows from \eqref{eq:CLplusisomorphism} that one has for the even sub-algebras
\begin{equation}
\mathcal{C}l_+(r,d-r, \mathbbm{R}) \cong \mathcal{C}l_+(d-r, r, \mathbbm{R}).
\end{equation}
In fact from \eqref{eq:CLplusisomorphism} this follows only for $r,d-r\geq 1$ but holds in fact also beyond this restriction.

The even sub-algebra $\mathcal{C}l_+(r,d-r, \mathbbm{R})$ has one generator less and also it's generators are different from the ones of the full algebra. Accordingly, also the matrices $\hat \gamma$ defined in \eqref{eq:defhatgamma} as well as $\beta$ and $\alpha$ defined in \eqref{eq:definitionBeta} and \eqref{eq:definitionBetaBar} are different. Nevertheless, these objects are still well defined and can be used for a classification of $\mathcal{C}l_+(r,d-r, \mathbbm{R})$, very similar as for the original algebra. Because the generators of the even sub-algebra can be taken as products of one particular generator $\gamma^{1-r}$ with all others, one can take the structure map in the even sub-algebra $\varsigma_\text{even}$ to be the reversal of the one particular direction $\gamma^{1-r} \to - \gamma^{1-r}$. 

\paragraph{Even Clifford sub-algebra for $d=1+3$.} Because of the importance for applications in physics, it is appropriate to discuss the reduction to the even sub-algebra in more detail on the example of Minkowski space $d=1+3$. For the full Clifford algebra $\mathcal{C}l(1,3,\mathbbm{R})$ of Minkowski space one has $\beta=\gamma^0$, $\alpha=\gamma^1\gamma^2\gamma^3$ and $\hat \gamma=\gamma^0\gamma^1 \gamma^2 \gamma^3$. For the even sub-algebra $\mathcal{C}l_+(1,3,\mathbbm{R})\cong\mathcal{C}l(0,3,\mathbbm{R})\cong \text{Mat}(2,\mathbbm{C})$ one has three ``space-like'' generators $\gamma^0 \gamma^1$, $\gamma^0 \gamma^2$ and $\gamma^0 \gamma^3$ and accordingly $\beta_\text{even}=\mathbbm{1}$, $\alpha_\text{even}=\hat \gamma_\text{even}=\hat \gamma = \gamma^0 \gamma^1 \gamma^2 \gamma^3$. The structure map is given by $\varsigma=\pm \hat \gamma$ in the full algebra, and $\hat \gamma$ commutes with all elements of the even sub-algebra and squares to minus one, $\hat \gamma^2=-\mathbbm{1}$. It provides a complex structure for the even sub-algebra. The structure map in the even sub-algebra $\varsigma_\text{even}$ corresponds (possibly up to an overall sign) to complex conjugation with respect to this complex structure. It is also implemented by the reversal $\gamma^0 \to -\gamma^0$.   

\paragraph{Even Clifford sub-algebra for $d=0+4$.} Let us also discuss the reduction to the even sub-algebra for Euclidean space $d=0+4$. For the full Clifford algebra $\mathcal{C}l(0,4,\mathbbm{R})$ one has $\beta=\mathbbm{1}$ and $\alpha=\hat\gamma=\gamma^1\gamma^2\gamma^3\gamma^4$. For the even sub-algebra $\mathcal{C}l_+(0,4,\mathbbm{R})\cong \mathcal{C}l(3,0,\mathbbm{R}) \cong \mathbbm{H} \oplus \mathbbm{H}$ one has three ``time-like'' generators $\gamma^1 \gamma^2$, $\gamma^1 \gamma^3$ and $\gamma^1 \gamma^4$ and accordingly $\beta_\text{even}=\hat\gamma_\text{even}=-\hat\gamma$ and $\alpha_\text{even}=\mathbbm{1}$. The structure map is given by $\varsigma=\pm \hat \gamma$ in the full algebra, and $\hat \gamma$ commutes with all elements of the even sub-algebra and squares to one, $\hat \gamma^2=\mathbbm{1}$. The sub-algebra has now the form of a direct sum $a\oplus b$ with $\hat \gamma = \pm \mathbbm{1}$ on the two summands. The structure map in the even sub-algebra $\varsigma_\text{even}$ corresponds (possibly up to an overall sign) to the interchange of the two direct summands $a\oplus b \to \pm b \oplus a$. It can be implemented as the reversal $\gamma^1 \to - \gamma^1$.

\begin{table}
\centering
\begin{tabular}{ c | c | c  }
\hline \hline
$r-(d-r) \text{ mod } 8$ & Isomorphic matrix algebra & Dimension  
\\ \hline
 1, 7 & $\text{Mat}(N, \mathbbm{R})$ & $N=2^{(d-1)/2}$
 \\ \hline
 3, 5 & $\text{Mat}(N, \mathbbm{H})$ & $N=2^{(d-3)/2}$ 
 \\ \hline
 2, 6 & $\text{Mat}(N, \mathbbm{C})$ & $N=2^{(d-2)/2}$
 \\ \hline
 4 & $\text{Mat}(N/2, \mathbbm{H}) \oplus \text{Mat}(N/2, \mathbbm{H})$ & $N=2^{(d-2)/2}$ 
 \\ \hline
 0 & $\text{Mat}(N/2, \mathbbm{R}) \oplus \text{Mat}(N/2, \mathbbm{R})$ & $N=2^{d/2}$
 \\ \hline \hline
\end{tabular}
\caption{Classification of even subalgebras of real Clifford algebras $\mathcal{C}l_+(r,d-r, \mathbbm{R})$ in terms of isomorphic real, complex or quaternionic matrix algebras.}
\label{tab:EvenCliffordAlgebraClassification}
\end{table}

\subsection{Canonical idempotents}

The inductive construction in the previous subsection has shown that the real Clifford algebras $\mathcal{C}l(r,d-r,\mathbbm{R})$ can be constructed as different tensor products of the basis elements $\mathcal{C}l(0,2,\mathbbm{R}) \cong \mathcal{C}l(1,1,\mathbbm{R}) \cong \text{Mat}(2,\mathbbm{R})$, $\mathcal{C}l(0,1,\mathbbm{R}) \cong \mathbbm{R} \oplus \mathbbm{R}$, $\mathcal{C}l(1,0,\mathbbm{R}) \cong \mathbbm{C}$ and $\mathcal{C}l(2,0,\mathbbm{R}) \cong \mathbbm{H}$. 

One can construct {\it canonical idempotents} by choosing in each tensor product factor $\text{Mat}(2,\mathbbm{R})$ one out of $p_{\pm} = (\mathbbm{1} \pm \sigma_3)/2$ and for each tensor product factor $\mathbbm{R} \oplus \mathbbm{R}$ one out of $p_+=(1,0)$ and $p_-=(0,1)$. For the remaining factors $\mathbbm{C}$ and $\mathbbm{H}$ one needs to chose the trivial idempotent $1$. In this way one obtains a {\it canonical set of commuting and non-annihilating idempotents}. Note that all these idempotents are hermitian in the sense of the definitions \eqref{eq:defHermitianConjugateGenerator} and \eqref{ABdagger}. 

We note that while this construction has been made with a view on the particular representation of the generators as a tensor product, it is not bound to this presentation. Because they correspond the a certain combinations of generators, the above idempotents can be constructed in any representation and will also be automatically hermitian, $\mathsf{H}(p)=p$.

\paragraph{Set of commuting and non-annihilating idempotents.} For a Clifford algebra with isomorphic $\mathbbm{R}$, $\mathbbm{C}$ or $\mathbbm{H}$ matrix representation of dimension $N=2^K$ one has $K$ non-trivial idempotents $p_1, \ldots, p_K$ that are mutually commuting and non-annihilating. In the construction of the isomorphic matrix algebra as a tensor product (in equation \eqref{eq:recursionCliffordAlgebras}), they correspond to canonical idempotents $p_\pm = (\mathbbm{1} \pm \sigma_3)/2$ on the different levels of the tensor product construction. For a Clifford algebra that is isomorphic to a direct sum $A \oplus B$ of such matrix algebras, there is also an additional idempotent $p_+=(1,0)$ or $p_-=(0,1)$ that is commuting and non-annihilating with all other idempotents. 

\paragraph{Canonical minimal idempotent.} We can pick one canonical idempotent for each matrix product factor, e.\ g.\ $(\mathbbm{1}+\sigma_3)/2$ for each tensor product factor $\text{Mat}(2,\mathbbm{R})$ and either $(1,0)$ or $(0,1)$ for a factor $\mathbbm{R} \oplus \mathbbm{R}$. This leads us to a {\it canonical minimal idempotent} $p$. In the canonical matrix representation as a tensor product it corresponds to a projector to a single coordinate direction. The canonical minimal idempotent is in particular hermitian in the sense that $p=\mathsf{H}(p)$. Such a hermitian minimal idempotent will be useful for the construction of spinor spaces further below.

\paragraph{Complete set of canonical idempotents and matrix representation.} 
By taking all kind of combinations of $p^{(j)}_{\pm}$ in the different tensor product factors, we obtain a set of canonical minimal idempotents $p_1, \ldots, p_N$ in the Clifford algebra that are orthogonal/annihilating and sum up to the unit element, $p_1+\ldots+p_N = \mathbbm{1}$. Such a set corresponds to a set of projectors to the different coordinates in the space of the tensor product matrix representation. A matrix element $M_{jk}\in \mathbbm{R}, \mathbbm{C}, \mathbbm{H}$ corresponding to an element $a$ of the Clifford algebra would be essentially given by $M_{jk} = p_j a p_k$, see ref.\ \cite{Vaz:2016qyw} for a more detailed discussion.

\subsection{Some examples up to five dimensions}

To gain some experience and intuition we will now discuss low dimensional real Clifford algebras in more explicit terms.

\paragraph{Three dimensions $d=0+3$.} For three spatial dimensions one has according to the construction above
\begin{equation}
\mathcal{C}l(0,3, \mathbbm{R}) \cong \mathcal{C}l(1,0, \mathbbm{R}) \otimes \mathcal{C}l(0,2, \mathbbm{R}) \cong \mathbbm{C} \otimes \text{Mat}(2, \mathbbm{R}) = \text{Mat}(2, \mathbbm{C}). 
\end{equation}
The generators are given by
\begin{equation}
\begin{split}
\gamma^1 =  i \otimes \begin{pmatrix} 0 & -1 \\ 1 & 0 \end{pmatrix} = \sigma_2, \quad\quad\quad
\gamma^2 =  1 \otimes \begin{pmatrix} 0 & 1 \\ 1 & 0 \end{pmatrix} = \sigma_1, \quad\quad\quad
\gamma^3 = 1 \otimes \begin{pmatrix} 1 & 0 \\ 0 & -1 \end{pmatrix} = \sigma_3.
\end{split}
\end{equation}
The product of all generators is given by $\hat\gamma=\gamma^1\gamma^2\gamma^3 = -i \mathbbm{1}$. 

Minimal idempotents are given by $1$ in the tensor product factor $\mathcal{C}l(1,0, \mathbbm{R}) \cong \mathbbm{C}$ times a convenient minimal idempotent in $\mathcal{C}l(0,2, \mathbbm{R}) \cong \text{Mat}(2, \mathbbm{R})$. For example, $(\mathbbm{1} \pm \gamma^3)/2$ are simple choices.

\paragraph{Three dimensions $d=3+0$.} In contrast, for three temporal dimensions one finds
\begin{equation}
\begin{split}
\mathcal{C}l(3,0, \mathbbm{R}) \cong \mathcal{C}l(0,1, \mathbbm{R}) \otimes \mathcal{C}l(2,0, \mathbbm{R}) \cong (\mathbbm{R} \oplus \mathbbm{R}) \otimes \mathbbm{H} = \mathbbm{H} \oplus \mathbbm{H}.
\end{split}
\end{equation}
Now the construction leads to
\begin{equation}
\begin{split}
\gamma^{-2} = & \gamma^1_{(0,1)} \otimes \begin{pmatrix} -i & 0 \\ 0 & i \end{pmatrix} = -i (1,-1)  \otimes \sigma_3, \\
\gamma^{-1} = & \mathbbm{1}_{(0,1)} \otimes \begin{pmatrix} 0 & -i \\ -i & 0 \end{pmatrix} = -i  (1,1) \otimes \sigma_1, \\
\gamma^0 = & \mathbbm{1}_{(0,1)}  \otimes \begin{pmatrix} 0 & -1 \\ 1 & 0 \end{pmatrix} = -i (1,1) \otimes \sigma_2.
\end{split}
\end{equation}
The product of all three generators is now 
\begin{equation}
\hat \gamma= \gamma^{-2}\gamma^{-1}\gamma^0=- \gamma^1_{(0,1)} = (-1,1)\otimes \mathbbm{1}.
\end{equation} 

In this case one can choose minimal idempotents in the factor $\mathcal{C}l(0,1, \mathbbm{R})\cong \mathbbm{R} \oplus \mathbbm{R}$ as either $(1,0)$ or $(0,1)$ while the only possibility in the factor $\mathcal{C}l(2,0, \mathbbm{R}) \cong \mathbbm{H}$ is the trivial idempotent $\mathbbm{1}$. In summary, the minimal idempotents are $(1,0)\otimes \mathbbm{1}$ and $(0,1)\otimes \mathbbm{1}$ while $(1,1) \otimes \mathbbm{1}$ is a non-minimal idempotent.

\paragraph{Four dimensions $d=0+4$ and $d=4+0$.} For the case of four space dimensions one finds
\begin{equation}
\mathcal{C}l(0,4, \mathbbm{R}) \cong \mathcal{C}l(2,0, \mathbbm{R}) \otimes \mathcal{C}l(0,2, \mathbbm{R}) \cong \mathbbm{H} \otimes \text{Mat}(2, \mathbbm{R}) = \text{Mat}(2, \mathbbm{H}). 
\end{equation}
The result is similar for four time directions,
\begin{equation}
\mathcal{C}l(4,0, \mathbbm{R}) \cong \mathcal{C}l(0,2, \mathbbm{R}) \otimes \mathcal{C}l(2,0, \mathbbm{R}) \cong \text{Mat}(2, \mathbbm{R}) \otimes \mathbbm{H} = \text{Mat}(2, \mathbbm{H}). 
\end{equation}
This is actually an example of a more general relation, there is a periodicity $\text{mod } 8$ in the signature $r-(d-r)$.

As an important case for physics applications, let us discuss the Euclidean algebra $\mathcal{C}l(0,4, \mathbbm{R})$ in more detail. A matrix representation of generators can be chosen as
\begin{equation}
\gamma^{1} = \begin{pmatrix} & - \mathbf{i} \\ \mathbf{i} & \end{pmatrix}, \quad\quad \gamma^{2} = \begin{pmatrix} & - \mathbf{j} \\ \mathbf{j} & \end{pmatrix}, \quad\quad
\gamma^3 = \begin{pmatrix} & 1 \\ 1 &  \end{pmatrix}, \quad\quad 
\gamma^4 = \begin{pmatrix} 1 & \\ & -1 \end{pmatrix}.
\end{equation}
where the entries are $\mathbbm{H}$ valued. Indeed, all these matrices square to unity and they anti-commute. The product of generators is given by
\begin{equation}
\hat \gamma = \gamma^1 \gamma^2 \gamma^3 \gamma^4 = \begin{pmatrix}  & \mathbf{k} \\ - \mathbf{k} & \end{pmatrix}.
\end{equation}
Idempotents can be chosen freely in the tensor product factor $\mathcal{C}l(0,2, \mathbbm{R}) \cong \text{Mat}(2, \mathbbm{R})$, for example $(\mathbbm{1}\pm \sigma_3)/2$, while for $\mathcal{C}l(2,0, \mathbbm{R})  \cong \mathbbm{H}$ one has only the trivial idempotent $\mathbbm{1}$. There is in particular the primitive idempotent
\begin{equation}
(\mathbbm{1} + \gamma^4) / 2= \begin{pmatrix} 1 & \\ & 0  \end{pmatrix}.
\end{equation}

\paragraph{Four dimensions $d=1+3$.} From the general classification given above one finds
\begin{equation}
\mathcal{C}l(1,3, \mathbbm{R}) \cong \mathcal{C}l(1,1, \mathbbm{R}) \otimes \mathcal{C}l(0,2, \mathbbm{R}) \cong \text{Mat}(2, \mathbbm{R}) \otimes \text{Mat}(2, \mathbbm{R}) \cong \text{Mat}(4, \mathbbm{R}). 
\end{equation}
Because this is the most important case for applications in physics (Minkowski space) we also discuss a concrete matrix representation. One may take as generators for example the set
\begin{equation}
\gamma^0 = \begin{pmatrix}  &  & -1 &  \\  &  &  & -1 \\ 1 & & &  \\ & 1 & & \end{pmatrix}, \quad
\gamma^1 = \begin{pmatrix}  & -1 &  &  \\ -1 &  &  & \\  & & & 1 \\ & & 1 & \end{pmatrix}, \quad 
\gamma^2 = \begin{pmatrix} -1 & &  &  \\ & 1 &  &  \\ & & 1 &  \\ & & & -1 \end{pmatrix}, \quad 
\gamma^3 = \begin{pmatrix}  & & 1 &  \\ &  &  & 1 \\ 1 & &  & \\ & 1 & &  \end{pmatrix}. 
\label{eq:gammad1p3}
\end{equation}
In this representation one has
\begin{equation}
\hat \gamma = \gamma^{0123}= \gamma^0 \gamma^1 \gamma^2 \gamma^3 = \begin{pmatrix}  & 1 & & \\ -1 & & & \\  & & & -1 \\ & & 1 & \end{pmatrix}.
\end{equation}

Idempotents can be chosen freely for both tensor product factors $\mathcal{C}l(1,1, \mathbbm{R}) \cong \text{Mat}(2, \mathbbm{R})$ and $\mathcal{C}l(0,2, \mathbbm{R}) \cong \text{Mat}(2, \mathbbm{R})$. Some examples are
\begin{equation}
\frac{1}{2}(\mathbbm{1} - \gamma^0 \gamma^3) = \begin{pmatrix} 1 & & & \\ & 1 & & \\ & & 0 & \\ & & & 0 \end{pmatrix}, \quad \frac{1}{2}(1-\gamma^2) = \begin{pmatrix} 1 & & & \\ & 0 & & \\ & & 0 & \\ & & & 1 \end{pmatrix}, \quad \frac{1}{2}(\mathbbm{1} - \gamma^0 \gamma^3) \frac{1}{2}(1-\gamma^2) = \begin{pmatrix} 1 & & & \\ & 0 & & \\ & & 0 & \\ & & & 0 \end{pmatrix}.
\label{eq:dempotd1p3}
\end{equation}
The last idempotent is primitive.

\paragraph{Four dimensions $d=3+1$.} This is a space with three time directions and one space direction or Minkowski space with an alternative choice of metric. Let us emphasize here that $\mathcal{C}l(1,3, \mathbbm{R})$ and $\mathcal{C}l(3,1, \mathbbm{R})$ are in fact {\it not} equivalent. Now the general construction gives
\begin{equation}
\mathcal{C}l(3,1, \mathbbm{R}) \cong \mathcal{C}l(1,1, \mathbbm{R}) \otimes \mathcal{C}l(2,0, \mathbbm{R}) \cong \text{Mat}(2, \mathbbm{R}) \otimes \mathbbm{H} \cong \text{Mat}(2, \mathbbm{H}). 
\end{equation}
Here one can work for example with the representation of generators
\begin{equation}
\gamma^{-2} = \begin{pmatrix} - \mathbf{i} & \\ & \mathbf{i} \end{pmatrix}, \quad\quad \gamma^{-1} = \begin{pmatrix} - \mathbf{j} & \\ & \mathbf{j} \end{pmatrix}, \quad\quad
\gamma^0 = \begin{pmatrix} & -1 \\ 1 &  \end{pmatrix}, \quad\quad \gamma^1 = \begin{pmatrix}  & 1 \\ 1 &  \end{pmatrix}.
\end{equation}
where the entries are $\mathbbm{H}$ valued. The product of generators gives
\begin{equation}
\hat \gamma = \gamma^{-2} \gamma^{-1} \gamma^0 \gamma^1 = \begin{pmatrix} -\mathbf{k} & \\ & \mathbf{k} \end{pmatrix}.
\end{equation}

The tensor product factor $\mathcal{C}l(1,1, \mathbbm{R}) \cong \text{Mat}(2, \mathbbm{R})$ has non-trivial idempotents, for example $(\mathbbm{1} \pm \sigma_3)/2$. In contrast for $\mathcal{C}l(2,0, \mathbbm{R}) \cong \mathbbm{H}$ one has only the trivial choice $\mathbbm{1}$. A primitive idempotent for $\mathcal{C}l(3,1, \mathbbm{R})$ is for example
\begin{equation}
(\mathbbm{1} - \gamma^0 \gamma^1)/2 = \begin{pmatrix} 1 & \\ & 0  \end{pmatrix}.
\end{equation}

\paragraph{Four dimensions $d=2+2$.} Finally, for two time and two space dimensions, the general classification given above yields
\begin{equation}
\mathcal{C}l(2,2, \mathbbm{R}) \cong \mathcal{C}l(1,1, \mathbbm{R}) \otimes \mathcal{C}l(1,1, \mathbbm{R}) \cong \text{Mat}(2, \mathbbm{R}) \otimes \text{Mat}(2, \mathbbm{R}) \cong \text{Mat}(4, \mathbbm{R}). 
\end{equation}
Idempotents can now again be freely chosen in both tensor product factors.

\paragraph{Five dimensions.} We also mention briefly the isomorphisms in five dimensions. One has
\begin{equation}
\begin{split}
\mathcal{C}l(0,5, \mathbbm{R}) & \cong \mathcal{C}l(0,1, \mathbbm{R}) \otimes \mathcal{C}l(2,0, \mathbbm{R}) \otimes \mathcal{C}l(0,2, \mathbbm{R}) 
\cong \text{Mat}(2,\mathbbm{H}) \oplus \text{Mat}(2,\mathbbm{H}), \\
\mathcal{C}l(1,4, \mathbbm{R}) & \cong  \mathcal{C}l(1,0, \mathbbm{R}) \otimes \mathcal{C}l(0,2, \mathbbm{R}) \otimes \mathcal{C}l(1,1, \mathbbm{R}) 
\cong \text{Mat}(4,\mathbbm{C}), \\
\mathcal{C}l(2,3, \mathbbm{R}) & \cong \mathcal{C}l(0,1, \mathbbm{R}) \otimes \mathcal{C}l(1,1, \mathbbm{R}) \otimes \mathcal{C}l(1,1, \mathbbm{R}) 
\cong \text{Mat}(4, \mathbbm{R}) \oplus \text{Mat}(4, \mathbbm{R}), \\
\mathcal{C}l(3,2, \mathbbm{R}) & \cong \mathcal{C}l(1,0, \mathbbm{R}) \otimes \mathcal{C}l(1,1, \mathbbm{R}) \otimes \mathcal{C}l(1,1, \mathbbm{R}) 
\cong \text{Mat}(4, \mathbbm{C}), \\
\mathcal{C}l(4,1, \mathbbm{R}) & \cong \mathcal{C}l(0,1, \mathbbm{R}) \otimes \mathcal{C}l(2,0, \mathbbm{R}) \otimes \mathcal{C}l(1,1, \mathbbm{R}) 
\cong \text{Mat}(2, \mathbbm{H}) \oplus \text{Mat}(2, \mathbbm{H}), \\
\mathcal{C}l(5,0, \mathbbm{R}) & \cong \mathcal{C}l(1,0, \mathbbm{R}) \otimes \mathcal{C}l(0,2, \mathbbm{R}) \otimes \mathcal{C}l(2,0, \mathbbm{R}) 
\cong \text{Mat}(4,\mathbbm{C}).
\end{split}
\end{equation}
It should now be clear how (minimal) idempotents for these cases can be constructed.

\section{Spinors}
\label{sec:Spinors}

Spinors are traditionally defined as quantities that transform with the {\it spin} group (or the larger {\it pin} group) in some matrix representation. There exists a more algebraic characterization in terms of {\it left and right ideals} of a Clifford algebra or one of the groups \cite{Lounesto:2001zz, Vaz:2016qyw} which is then actually formally independent of a particular matrix representation. This construction will be reviewed here after some algebraic preliminaries. The connection to the matrix representation will also become clear.

\subsection{Spinor spaces as minimal or quasi-minimal ideals}
For the following, the relevant Clifford algebra could be the full real algebra $\mathcal{C}l(r,d-r,\mathbbm{R})$ or the even subalgebra $\mathcal{C}l_+(r,d-r,\mathbbm{R})$. The construction could also be based on a complex Clifford algebra. In the following we we will keep this somewhat open and just refer to some Clifford algebra $\mathcal{C}l$.
 
\paragraph{Ideals.} A subset $\mathcal{L}$ of the Clifford algebra $\mathcal{C}l$ is called {\it left ideal} of $\mathcal{C}l$ if for all $a\in \mathcal{C}l$ and $\psi \in \mathcal{L}$ one has
\begin{equation}
a \psi \in \mathcal{L}.
\end{equation}
In an analogous way, a subset $\mathcal{R}$ is called {\it right ideal} of $\mathcal{C}l$ if for all $a\in \mathcal{C}l$ and $\chi \in \mathcal{R}$ one has
\begin{equation}
\chi a \in \mathcal{R}.
\end{equation}
Finally, a subset $\mathcal{I}$ of the Clifford algebra $\mathcal{C}l$ is called {\it bilateral ideal} or simply {\it ideal} of $\mathcal{C}l$ if for all $a,b\in \mathcal{C}l$ and $\Psi \in \mathcal{I}$ one has
\begin{equation}
a \Psi b \in \mathcal{I}.
\end{equation}
All these ideals are vector spaces. 

For a matrix algebra, an example of a left ideal are matrices with only one or several columns non-vanishing, e.\ g.\
\begin{equation}
M= \begin{pmatrix} a_1 & 0  \\ a_2 & 0  \end{pmatrix}.
\label{eq:matrixIdeal}
\end{equation}
The subset of matrices of this form is mapped to itself under matrix multiplication. In a similar way, a matrix with only one non-vanishing row is an example of a right ideal. 

\paragraph{Minimal ideal.} A {\it minimal ideal} is an ideal that does not contain any non-trivial {\it subideals}, i.\ e.\ the only subideals are the zero element and the ideal itself. The matrix set in \eqref{eq:matrixIdeal} is an example for a minimal left ideal. 

In passing we note that minimal ideals are also useful for a decomposition of representations of Clifford algebra into irreducibles:

\paragraph{Regular representation of Clifford algebra.} The Clifford algebra $\mathcal{C}l$ has a representation in terms of maps (endomorphisms) on itself. To every element $a\in\mathcal{C}l$ one can associate the map $L(a)$ such that for $X\in \mathcal{C}l$ one has
\begin{equation}
L(a) X = a X.
\end{equation}
Obviously $L(ab)=L(a) L(b)$. This is called the {\it regular representation.} One may also define $R(a)$ such that
\begin{equation}
R(a) X = X a.
\end{equation}
In this case $R(ab) = R(b) R(a)$, so this is a representation of the {\it reversed} or {\it opposed} algebra.

\paragraph{Decomposition into irreducibles.}
One can decompose any representation of a Clifford algebra into sums of irreducible representations which are then by definition minimal ideals.

The definition of spinors we will use further below is based on minimal left or right ideals. There is a particularly convenient method to construct such minimal ideals in terms of  idempotents.

\paragraph{Ideal from projector or idempotent.} Given a projector or idempotent $p$ one can define a left ideal $\mathcal{L}$ of the Clifford algebra $\mathcal{C}l$ as the set of elements $a p$ with $ a \in \mathcal{C}l$. In other words $\mathcal{L}=\mathcal{C}l \; p$. In a similar way, a right ideal is given by $\mathcal{R} = p \; \mathcal{C}l$. If the projector or idempotent is {\it primitive}, the resulting ideal is a {\it minimal ideal}. This is argueably the simplest way to construct a minimal ideal.

We now define spinor spaces in terms of ideal. It is sometimes convenient to work with spinors as featuring irreducible representations of the algebra, in which case the corresponding spinor space is a minimal ideal, but it is also sometimes convenient to relax this and to work in reducible spinor spaces corresponding to non-minimal ideals.

\paragraph{Space of column spinors as (minimal) left ideal.} One may define the space $r$ of column spinors (with the name referring to the matrix representation) of the Clifford algebra $\mathcal{C}l$ as a (minimal) left ideal $\mathcal{C}l \, p$ where $p$ is a (primitive) canonical idempotent. By taking a canonical idempotent we make sure that $p$ is hermitian, $p = \mathsf{H}(p)$. It is clear that the space $r$ constitutes a {\it representation of the Clifford algebra} $\mathcal{C}l$. Moreover, one can multiply these spinors from the right by elements of the division ring $p\, \mathcal{C}l \, p$ without leaving the spinor space $r$.

\paragraph{Space of row spinors as a (minimal) right ideal.} In a very similar way one can define the space of row spinors as corresponding to a (minimal) right ideal $p^\prime \mathcal{C}l$ where $p^\prime$ is again a (primitive) idempotent. Such row spinors can be multiplied from the left by elements of the division ring $p^\prime \, \mathcal{C}l \, p^\prime$. Moreover, we for canonical idempotent $p$ we can take $p^\prime=\mathsf{H}(p) = p$, such that the division rings $p^\prime \, \mathcal{C}l \, p^\prime$ and $p \, \mathcal{C}l \, p$ actually agree.

\paragraph{Transformations in spinor space: adjoint (naive) implementation.} In our formulation, spinors are themselves part of the Clifford algebra and could therefore also transform under various maps within the Clifford algebra such as e.\ g.\ grade involution. Some care is needed here, however, as will be discussed below. We first discuss a naive implementation and subsequently a more proper one. 

To a given spinor $\psi$ out of a (minimal) left ideal $r=\mathcal{C}l \, p$ one may define various involutions or other transformations as for any element of the Clifford algebra. For example, the grade involution of $\psi$ could be defined by $G(\psi) = \varsigma \psi \varsigma^{-1}$. It would be part of the left ideal $\mathcal{C}l \, \mathsf{G}(p)$ featuring a regular representation of the Clifford algebra $\mathcal{C}l$. It is important to note, however, that in general $\mathsf{G}(p) = \varsigma p \varsigma^{-1} \neq p$ such that the left ideals $\mathcal{C}l \, \mathsf{G}(p)$ and $\mathcal{C}l \, p$ do {\it not agree}. In other words, this leads to the problem that a spinor $\psi$ and its grade involution $\mathsf{G}(\psi)$ are not in the same spinor space. The problem appears also for other transformations, such as the action of the pin group. In the following we discuss how this problem can be solved so that spinors and transformed spinors are elements of the same left ideal.

\paragraph{Transformations in spinor space: regular (proper) implementation.} To circumvent the problem described above, one has to work with a different implementation of transformations for spinors. As an exemplary  and rather important map in the Clifford algebra let us consider again the grade involution $\mathsf{G}$. In terms of the structure map \eqref{eq:StructureMapCliffordAlgebra}, one can decompose grade involution into a left and a right action, $\mathsf{G}(a) = \varsigma a \varsigma^{-1}$. If one takes only the left action to transform the spinor $\psi$ it has a chance to be part of the same left ideal. Specifically, for a given spinor $\psi \in S = \mathcal{C}l \, p$, we use instead of grade involution the left action of the structure map and set 
\begin{equation}
\mathsf{G}(\psi) = \varsigma \psi.
\label{eq:StructureMapSpinorFundamental}
\end{equation}
For a spinor $\psi = a p $ with $a \in \mathcal{C}l$ and primitive idempotent $p$ we have now
\begin{equation}
\varsigma \psi = \varsigma a \varsigma^{-1} \varsigma p = \mathsf{G}(a) \varsigma p
\end{equation}
This shows that in cases where the structure map is itself part of the Clifford algebra, $\varsigma \in \mathcal{C}l$ or when $\varsigma p = p$, the spinor spaces or left ideals $\mathcal{C}l p$ and $\varsigma\mathcal{C}l p$ actually agree. 

Let us recall here that $\varsigma$ as defined through eq.\ \eqref{eq:StructureMapCliffordAlgebra} and $\varsigma^2=\pm \mathbbm{1}$ is unique only up to a sign. Accordingly, also the transformation on a spinor \eqref{eq:StructureMapSpinorFundamental} is only unique up to a sign. However, no ambiguity arises for expressions that involve an even number of spinors. 

One might be tempted to demand a transformation behavior similar to \eqref{eq:StructureMapSpinorFundamental}  also for row spinors $\chi \in p \mathcal{C}l$. If they would transform under the right action of the structure map like $\mathsf{G}(\chi) = \chi \varsigma^{-1}$ we would find that a product of a row and a column spinor is invariant $\mathsf{G}(\chi \psi) = \chi \psi$. In other words, such products would always be part of the even sub-algebra. It turns out that this is in general too restrictive. We will later connect row spinors through certain conjugation relations with column spinors. They inherit their transformation behavior under the structure map or other transformations in a natural way through these relations.

\paragraph{Minimal spinor spaces.} For $d$ even, the structure map $\varsigma=\pm \hat \gamma$ is itself part of the Clifford algebra so that $\varsigma \mathcal{C} l \, p = \mathcal{C}l \, p$ is guaranteed. In other words, a spinor $\psi$ and its grade involution $\varsigma \psi$ are indeed in the same space.

For $d$ odd, the situation is slightly more complicated. As discussed below \eqref{eq:StructureMapCliffordAlgebra}, for $r-(d-r)=1, 5 \text{ mod }8$, the structure map is a complex conjugation. Because $p^2=p$, an idempotent must be real and is therefore unchanged by this complex conjugation. In other words, $\varsigma p = p$ and therefore $\varsigma \mathcal{C}l \, p = \mathcal{C} l \, p$ so that again a spinor and its grade involution live in the same space.

In contrast, for $r-(d-r)=3, 7 \text{ mod }8$, the algebra has the structure of a direct sum $A \oplus B$; it is reducible. The structure map interchanges the two summands. If now $p$ is a {\it minimal} or primitive idempotent which generates a {\it minimal} left ideal, it can only be non-zero in one of these two summands. In other words, a primitive idempotent would contain a tensor product factor $(1,0)$ or $(0,1)$. In this case $\varsigma p \neq p$ and a spinor $\psi \in \mathcal{C}l \, p$ and its grade involute $\varsigma \psi \in \mathcal{C}l \, \varsigma \, p$ are {\it not} in the same space. One could also formulate this as the statement that for $r-(d-r)=3, 7 \text{ mod }8$ a minimal left ideal is annihilated by grade involution $\mathsf{G}$ or the structure map $\varsigma$.

\paragraph{Quasi-minimal spinor spaces.} However, $p$ can also be reducible and contain a tensor product factor $(1,1)$. In that case $\varsigma p = p$ and spinors and their grade involutes live indeed in the same space, $\mathcal{C}l \, p = \mathcal{C}l \, \varsigma \, p$. Such non-minimal spinor spaces are reducible in the sense of standard representation theory of Clifford algebras. However, they are nevertheless minimal in the sense that they are the smallest possible spinor space that features a non-trivial representation of the entire Clifford algebra and is not annihilated by grade involution $\mathsf{G}$. We will call such representations {\it quasi-minimal}. They are generated by quasi-minimal idempotents as they have been defined in section \ref{sec:Idempotents}.

\paragraph{Time reversed and space reversed spinors.} A similar construction works also for the time reversal and space reversal. We define, based on \eqref{eq:timeReversalTransformation} for a given spinor $\psi\in \mathcal{C}l\, p$ the time reversed spinor
\begin{equation}
\mathcal{T}(\psi) = \beta \varsigma^r \psi.
\label{eq:TimeReflectionSpinor}
\end{equation}
Note that for a real Clifford algebra $\beta$ (similar to $\varsigma$) is defined uniquely only up to a sign. In a similar way we define based on \eqref{eq:SpaceReversalTransformation} the space reversed spinor as
\begin{equation}
\mathcal{P}(\psi) = \alpha \varsigma^{d-r} \psi.
\label{eq:SpaceReflectionSpinor}
\end{equation}
Again this is unique up to a sign for a real Clifford algebra.

We note here that for $d$ odd, one of the transformations \eqref{eq:TimeReflectionSpinor} or \eqref{eq:SpaceReflectionSpinor} includes a structure map $\varsigma$. In order to be {\it not} annihilated by time or space reversal transformations when $r-(d-r)=3,7 \text{ mod }8$ one needs the primitive idempotent $p$ that generates the spinor space as a left ideal to contain a tensor product factor $(1,1)$ and to be accordingly non-minimal. However, it can be {\it quasi-minimal} in the sense defined above. On the other side, spinors as part of a {\it minimal} left ideal are necessarily annihilated by either time reversal or space reversal, i.e. they are not invariant under one of these discrete symmetries.

\paragraph{Action of pin group on spinors.} Let us recall that the action of the pin group on a general element of the Clifford algebra can be written as in equation \eqref{eq:actionPinGroupCliffordAlgebraElement}. One may now ask how one can define the action of pin group elements on spinors such that spinor spaces are invariant. A subtlety arises here from the fact that, even after normalization, pin group elements are actually only unique up to a sign. Indeed, the transformation  \eqref{eq:actionPinGroupCliffordAlgebraElement} is the same when $a\to -a$. If the pin group element $a$ can be continuously deformed to the unit element $\mathbbm{1}$ one can fix it's sign, but more generally, there is no possibility to do this. 

Related to this is the following problem. Based on eq.\ \eqref{eq:actionPinGroupCliffordAlgebraElement} one may define the action of the pin group on a spinor as
\begin{equation}
\psi \to a \varsigma^{\mathsf{g}(a)} \psi.
\label{eq:pinGroupOnSpinors}
\end{equation}
Performing two such transformations gives
\begin{equation}
\psi \to b \varsigma^{\mathsf{g}(b)} a \varsigma^{\mathsf{g}(a)} \psi = (-1)^{\mathsf{g}(a) \cdot \mathsf{g}(b)} (ba) \varsigma^{\mathsf{g}(ba)}  \psi .
\end{equation}
This shows that \eqref{eq:pinGroupOnSpinors} is in fact only a representation of the pin group up to the overall sign. In other words, it is a proper representation when $\psi$ and $-\psi$ are identified. In practice, the sign ambiguity is not very severe. Physical observables are actually spinor bilinears of some type and if both spinors are transformed consistently, the overall sign drops out again.

\paragraph{Action of spin group on spinors.}
The action of the spin group on a column spinor is obtained by specializing eq.\ \eqref{eq:pinGroupOnSpinors} to even elements $a\in \text{Spin}(r,d-r,\mathbbm{R}) = \text{Pin}(r,d-r,\mathbbm{R}) \cap \mathcal{C}l_+(r,d-r,\mathbbm{R})$,
\begin{equation}
\psi \to a \psi.
\label{eq:spinGroupOnSpinors}
\end{equation}
Note that in contrast to the pin group, no sign ambiguity appears here.

\paragraph{Remark on quaternionic spinors.} The fact that quaternion-valued spinors are necessary in certain dimensions if one works in the framework of a real Clifford algebra may come as a surprise. Let us discuss the simplest incarnation of such spinors for the algebra $\mathcal{C}l(2,0,\mathbbm{R})$ in more detail here. A matrix representation has been given in \eqref{eq:generators20a} and \eqref{eq:generators20b} and in that representation every element of the Clifford algebra can be written as
\begin{equation}
a \mathbbm{1} + b \gamma^{-1} + c \gamma^0 +d \hat\gamma = \begin{pmatrix} a -i d, & -i b - c\\ -i b+c, & a +i d \end{pmatrix}.
\end{equation}   
One observes that these are complex two-by-two matrices. In the traditional approach one would work with spinors as column vectors with two complex entries. However, one can write every such column vector as
\begin{equation}
\begin{pmatrix} a-id \\ c-i b \end{pmatrix} = \begin{pmatrix} a -i d, & -i b - c\\ -i b+c, & a +i d \end{pmatrix} \begin{pmatrix} 1 \\ 0 \end{pmatrix}.
\label{eq:complexQuanternionicColumnSpinor}
\end{equation}   
In other words, one has here not only a map from the Clifford algebra to it's spinor representation but also a map from the spinor back to the entire Clifford algebra element. This shows in which sense the spinors can here actually be understood as elements of the Clifford algebra $\mathcal{C}l(2,0,\mathbbm{R}) \cong \mathbbm{H}$ itself and do {\it not} form a non-trivial ideal.

\paragraph{Quaternionic structure and symplectic Majorana spinors.} Seen as a complex vector space, the spinors $\psi$ in \eqref{eq:complexQuanternionicColumnSpinor} have a {\it quaternionic structure}. This is an anti-linear map $J$ such that
\begin{equation}
J (\lambda \psi) = \lambda^* J(\psi), \quad\quad\quad\text{for} \quad \lambda \in \mathbbm{C},
\end{equation}
and with $J(J(\psi))=-\psi$.  Concretely, this map $J$ is here given for example by complex conjugation together with multiplication by the anti-symmetric, real matrix $\gamma^0$ in \eqref{eq:generators20a}. 

If one can now somehow define another map $K$ on the space of spinors (typically using additional structure such as a flavor index or possibly employing the reversal of a coordinate \cite{Wetterich:2010ni}) with the property $K K^*=- \mathbbm{1}$ one can define {\it symplectic Majorana spinors} by the condition
\begin{equation}
K J \psi = \pm \psi.
\label{eq:symplecticMajoranaSpinorCondition}
\end{equation}
This is consistent in the sense that $KJKJ=K K^* J^2=(-\mathbbm{1})^2=\mathbbm{1}$. With the condition \eqref{eq:symplecticMajoranaSpinorCondition} the number of real dimensions is typically reduced by a factor $2$. We emphasize again that additional structure is needed for this construction.

\subsection{Conjugate spinors}

\paragraph{Conjugate or adjoint spinors.} To a given space of column spinors $\mathcal{C}l\, p$ generated by the (minimal or primitive) canonical idempotent $p$ one can relate a spaces of row spinors or conjugate spinors in several ways. In fact, $p\, \mathcal{C}l$, $\mathsf{C}(p)\, \mathcal{C}l$ and $\mathsf{R}(p)\, \mathcal{C}l$ would be examples for such row spinor spaces. The question arises here what is the most useful starting point for this construction, or in other words, what is the most convenient way to associate a row spinor to a given column spinor as elements of the Clifford algebra. We note here in particular that a notion of transpose is not directly available in the Clifford algebra. We do have the notion of the hermitian conjugate, eq.\ \eqref{eq:HermitianConjugationGeneral} available within the Clifford algebra, and this provides indeed the most useful starting point.

Let us associate to a given spinor space $S=\mathcal{C}l\, p$ generated by the (minimal or primitive) canonical idempotent $p$ the space of conjugate spinors $\mathsf{H}(S) = \mathsf{H}(p) \mathcal{C}l = p \, \mathcal{C}l $. We have used here that a canonical idempotent is hermitian, $\mathsf{H}(p) = p$. This is now a (minimal) right ideal, generated by the idempotent $\mathsf{H}(p) = p$. Moreover, one can in fact associate to every spinor $\psi = a p$ its {\it hermitian conjugate spinor}
\begin{equation}
\psi^\dagger = \mathsf{H}(\psi) = p \mathsf{H}(a).
\end{equation}
The conjugate spinor space $\mathsf{H}(S)$ is also invariant under multiplication from the left by elements of the division algebra $p \mathcal{C}l p$ which is isomorphic to $\mathbbm{R}$, $\mathbbm{C}$ or $\mathbbm{H}$. 

\paragraph{Dirac adjoint spinors.} On this basis we define now various other row spinors. The first Dirac adjoint can be defined as
\begin{equation}
\mathsf{D}_1(\psi) = \mathsf{H}(\psi) \alpha \varsigma^{d-r}  = \alpha \varsigma^{d-r} \mathsf{C}(\psi). 
\label{eq:firstDiracAdjointSpinor}
\end{equation}
In a similar way, the second Dirac adjoint is given by
\begin{equation}
\mathsf{D}_2(\psi) = \mathsf{H}(\psi) \beta \varsigma^{r}  = \beta \varsigma^{r} \mathsf{R}(\psi). 
\label{eq:secondDiracAdjointSpinor}
\end{equation}
We have employed here only the right action part of the transformations \eqref{eq:Dirac1GeneralClifford} and \eqref{eq:Dirac2GeneralClifford}, respectively. This makes sure that $\mathsf{H}(\psi)$, $\mathsf{D}_1(\psi)$ and $\mathsf{D}_2(\psi)$ are indeed part of the same row spinor space or right ideal $\mathsf{H}(S)$. 

Note that one can relate the first and second Dirac adjoint as 
\begin{equation}
\mathsf{D}_2(\psi) = \mathsf{D}_1(\psi)   \varsigma^{-(d-r)} \alpha^{-1} \beta \varsigma^r.
\end{equation}
In particular, for $d$ odd this relation includes a power of the structure map $\varsigma$. For the complex Clifford algebras that occur when $r-(d-r)=1,5 \text{ mod }8$, the structure map $\varsigma$ contains a complex conjugation and the two Dirac adjoints are in this sense complex conjugates of each other. For $r-(d-r)=3,7 \text{ mod }8$ the structure map interchanges the two summands in the reducible tensor product factor of the Clifford algebra. In contrast to this, when $d$ is even, the first and second Dirac adjoint differ simply by a factor within the Clifford algebra, namely $\pm \hat \gamma$. (In fact, this agrees with the structure map when $d$ is even.) We see here that one of the two ``Dirac adjoint spinors'' could in fact have been called a ``Majorana adjoint spinor''.

As an example, in $d=4$ dimensions one has in Minkowski space with $r=1$ and $\varsigma=\hat \gamma$ that $\mathsf{D}_1(\psi) = \psi^\dagger \gamma^0$ (the standard Dirac adjoint in Minkowski space) and $\mathsf{D}_2(\psi) = -\psi^\dagger \gamma^0 \hat \gamma$. In contrast, for the Euclidean case $r=0$ one finds $\mathsf{D}_1(\psi) = \psi^\dagger \hat \gamma$ and $\mathsf{D}_2(\psi) = \psi^\dagger$. We have used here the transformations in the full algebra $\mathcal{C}l(1,3,\mathbbm{R})$. In this sense $\psi$ is a representation of the pin group or a {\it pinor}. There is a similar definition of Dirac adjoints in the {\it even} sub-algebra $\mathcal{C}l_+(1,3,\mathbbm{R}) \cong \mathcal{C}l(3,0,\mathbbm{R})$. In that case $\psi$ is a representation of the spin group or a {\it proper spinor}. In that case, $\mathsf{D}_1(\psi)$ and $\mathsf{D}_2(\psi)$ would differ essentially by a complex conjugation, see the discussion at the end of section \ref{sec:InductiveConstruction}.

\paragraph{Action of pin group on Dirac adjoint spinors.} By using the transformation law \eqref{eq:pinGroupOnSpinors} in the definitions \eqref{eq:firstDiracAdjointSpinor} and \eqref{eq:secondDiracAdjointSpinor} one finds the transformation behavior of the two Dirac adjoints under the pin group $a\in \text{Pin}(r,d-r,\mathbbm{R})$,
\begin{equation}
\begin{split}
\mathsf{D}_1(\psi) & \to \mathsf{D}_1(a \varsigma^{\mathsf{g}(a)}\psi) = (-1)^{(d-r)\mathsf{g}(a)} \, \mathsf{D}_1(\psi) \, (\varsigma^{-1})^{\mathsf{g}(a)} \, \mathsf{C}(a), \\
\mathsf{D}_2(\psi) & \to \mathsf{D}_2(a \varsigma^{\mathsf{g}(a)}\psi) = (-1)^{r\mathsf{g}(a)} \, \mathsf{D}_2(\psi) \, (\varsigma^{-1})^{\mathsf{g}(a)} \, \mathsf{R}(a).
\end{split}
\label{eq:actionPinGroupDiracAdjoints}
\end{equation}
This rather general transformation behavior can now be specialized, for example to even elements of the spin group $a$ with $\mathsf{g}(a)=0$. The transformation behavior becomes for $a\in \text{Spin}(r,d-r,\mathbbm{R})$ simply
\begin{equation}
\begin{split}
\mathsf{D}_1(\psi) & \to \mathsf{D}_1(a \psi) = \mathsf{D}_1(\psi) \, \mathsf{C}(a), \\
\mathsf{D}_2(\psi) & \to \mathsf{D}_2(a \psi) = \mathsf{D}_2(\psi) \, \mathsf{R}(a).
\end{split}
\end{equation}
Note that for the even elements in the spin group one has $\mathsf{C}(a) = \mathsf{R}(a)$. One can also specialize to the discrete transformations of time and space reversal.

\paragraph{Time and space reversals of Dirac adjoint spinors.} One can work out how the first and second Dirac adjoint spinors \eqref{eq:firstDiracAdjointSpinor} and \eqref{eq:secondDiracAdjointSpinor} transform under time and space reversal by taking the corresponding Dirac adjoints of the time and space reversed spinors in eqs.\ \eqref{eq:TimeReflectionSpinor} and \eqref{eq:SpaceReflectionSpinor}. One can in fact consider this as a special case of \eqref{eq:actionPinGroupDiracAdjoints}. One finds for time reversal
\begin{equation}
\begin{split}
\mathsf{D}_1(\mathcal{T}(\psi)) & = \mathsf{D}_1(\beta \varsigma^{r} \psi) = (-1)^{(d-r)r}\mathsf{D}_1(\psi) (\varsigma^{-1})^r \mathsf{C}(\beta) = (-1)^{(d-r)r}\mathsf{D}_1(\psi) (\varsigma^{-1})^r \beta^{-1}, \\
\mathsf{D}_2(\mathcal{T}(\psi)) & = \mathsf{D}_2(\beta \varsigma^{r} \psi) = (-1)^{r}\mathsf{D}_2(\psi) (\varsigma^{-1})^r \mathsf{R}(\beta) = \mathsf{D}_2(\psi) (\varsigma^{-1})^r \beta^{-1},
\end{split}
\end{equation}
and similar for space reversal,
\begin{equation}
\begin{split}
\mathsf{D}_1(\mathcal{P}(\psi)) & = \mathsf{D}_1(\alpha \varsigma^{d-r} \psi) = (-1)^{(d-r)}\mathsf{D}_1(\psi) (\varsigma^{-1})^{d-r} \mathsf{C}(\alpha) = \mathsf{D}_1(\psi) (\varsigma^{-1})^{d-r} \alpha^{-1}, \\
\mathsf{D}_2(\mathcal{P}(\psi)) & = \mathsf{D}_2(\alpha \varsigma^{d-r} \psi) = (-1)^{r(d-r)}\mathsf{D}_2(\psi) (\varsigma^{-1})^{d-r} \mathsf{R}(\alpha) = (-1)^{r(d-r)} \mathsf{D}_2(\psi) (\varsigma^{-1})^{d-r} \alpha^{-1}.
\end{split}
\end{equation}

\paragraph{Spinor inner product.} 
For a given column spinor $\psi \in S = \mathcal{C}l \, p$ and row spinors $\chi \in \mathsf{H}(S) = p \, \mathcal{C}l$ we can consider the product
\begin{equation}
\chi \psi \in p \, \mathcal{C}l \, p.
\end{equation}
As stated in section \ref{sec:Idempotents}, the subset $p\, \mathcal{C}l \, p$ is in fact a division ring and isomorphic to $\mathbbm{R}$, $\mathbbm{C}$ or $\mathbbm{H}$ when $p$ is an idempotent. Which of these three cases appears is of course directly linked to the matrix algebra isomorphic to $\mathcal{C}l$.

\paragraph{First and second inner product of column spinors.} On this basis, we can define two inner products of column spinors $\psi, \varphi \in S = \mathcal{C}l\, p$. They are given by
\begin{equation}
\mathsf{D}_1(\varphi) \psi = \mathsf{H}(\varphi)  \alpha \varsigma^{d-r} \psi = \alpha \varsigma^{d-r} \mathsf{C}(\varphi) \psi ,
\label{eq:InnerProductSpinor1}
\end{equation}
and
\begin{equation}
\mathsf{D}_2(\varphi) \psi = \mathsf{H}(\varphi)  \beta \varsigma^{r} \psi = \beta \varsigma^{r} \mathsf{R}(\varphi) \psi,
\label{eq:InnerProductSpinor2}
\end{equation}
respectively, and are both part of $p\, \mathcal{C}l \, p$. Let us emphasize again that these inner products of spinors in a real Clifford algebra are {\it not} necessarily real but in general either an element of the real numbers $\mathbbm{R}$, the complex numbers $\mathbbm{C}$, or the quaternions $\mathbbm{H}$. 

It is immediately clear that the two inner products \eqref{eq:InnerProductSpinor1} and \eqref{eq:InnerProductSpinor2} are invariant under the action \eqref{eq:spinGroupOnSpinors} of the restricted spin group \eqref{eq:restrictedSpinGroup}. In fact, for $\psi \to a \psi$ and $\varphi \to a \varphi$ we have
\begin{equation}
\mathsf{C}(\varphi) \psi \to \mathsf{C}(\varphi)  \mathsf{C}(a) a\psi = \mathsf{C}(\varphi) \psi, \quad\quad\quad 
\mathsf{R}(\varphi) \psi \to \mathsf{R}(\varphi)  \mathsf{R}(a) a\psi = \mathsf{R}(\varphi) \psi,
\end{equation}
as a consequence of eq.\ \eqref{eq:restrictedSpinGroup}.

\subsection{Pinors}

We discuss now spinor spaces that feature non-trivial representations of the entire (i.\ e.\ not only even) real Clifford algebra $\mathcal{C}l(r,d-r,\mathbbm{R})$. Because they feature a representation of the pin group $\text{Pin}(r,d-r,\mathbbm{R})$, such spinors are sometimes called {\it pinors}.

We take the spinor space $S$ to be a {\it minimal} or {\it quasi-minimal} left ideal $\mathcal{C}l(r,d-r,\mathbbm{R}) p$ where $p$ is a {\it minimal} or {\it quasi-minimal} idempotent. As we have remarked above, a minimal left ideal is necessarily annihilated by the structure map $\varsigma$ when $r-(d-r)=3,7 \text{ mod }8$ and in that case it is sometimes useful to work with a {\it quasi-minimal} left ideal instead.

In the following we go through the different cases where the real Clifford algebra is isomorphic to different matrix algebras. For this discussion it is useful to keep the classification in table \ref{tab:CliffordAlgebraClassification} in mind. In order to have a complete characterization for future reference we accept that the discussion in this and the subsequent subsection is partly repetitive.

\subsubsection{Cases with $r-(d-r)=0,6 \text{ mod }8$}\label{sec:pinor06}
Examples in up to six dimensions are $d=0+2$, $d=1+1$, $d=1+3$ (Minkowski space), $d=2+2$, $d=2+4$, $d=3+3$, $d=6+0$. 

Here the real Clifford algebra $\mathcal{C}l(r,d-r,\mathbbm{R})$ is isomorphic to a {\it real} matrix algebra $\text{Mat}(N, \mathbbm{R})$ with $N=2^{d/2}$ dimensions. Accordingly, column pinors are isomorphic to a column vector with $2^{d/2}$ {\it real} entries. 

The Clifford structure map is given by $\varsigma=\pm \hat \gamma$ where the ``volume element'' $\hat \gamma$ as defined in \eqref{eq:defhatgamma} and the elements for time reversal $\beta$ (defined in \eqref{eq:definitionBeta}) and space reversal $\alpha$ (defined in \eqref{eq:definitionBetaBar}) can all be represented by {\it real} matrices. Both the time reversed pionor defined in \eqref{eq:TimeReflectionSpinor} and the space reversed pinor defined in \eqref{eq:TimeReflectionSpinor} are part of the original pinor space $S$.

The two Dirac adjoints of a column {\it pinor} as defined in \eqref{eq:firstDiracAdjointSpinor} and \eqref{eq:secondDiracAdjointSpinor} are now isomorphic to {\it real} row vectors with $2^{d/2}$ entries. They differ essentially by a factor $\hat \gamma$. 

Column and row pinors can be multiplied with real numbers to yield another such pinor. The spinor inner products defined in \eqref{eq:InnerProductSpinor1} and \eqref{eq:InnerProductSpinor2} yield both real numbers. 

Relativistic fermions described by these pinors correspond to {\it Majorana fermions}.

\subsubsection{Cases with $r-(d-r)=2,4 \text{ mod }8$}\label{sec:pinor24}
Examples in up to six dimensions are $d=2+0$, $d=0+4$ (Euclidean space), $d=3+1$ (Minkowski space with ``mainly minus'' metric), $d=4+0$, $d=0+6$, $d=1+5$, $d=4+2$, $d=5+1$. 

Here the real Clifford algebra $\mathcal{C}l(r,d-r,\mathbbm{R})$ is isomorphic to a {\it quaternionic} matrix algebra $\text{Mat}(N, \mathbbm{H})$ with $N=2^{(d-2)/2}$ dimensions. Accordingly, column pinors are isomorphic to a column vector with $2^{(d-2)/2}$ {\it quaternionic} entries. Note that this corresponds to a real dimension $2^{(d+2)/2}$ which is a factor $2$ more than for the real case $r-(d-r)=0,6 \text{ mod }8$ with the same dimension $d$. 

The Clifford structure map is given by $\varsigma=\pm \hat \gamma$ where the ``volume element'' $\hat \gamma$ as defined in \eqref{eq:defhatgamma} and the elements for time reversal $\beta$ (defined in \eqref{eq:definitionBeta}) and space reversal $\alpha$ (defined in \eqref{eq:definitionBetaBar}) can all be represented by {\it quaternionic} valued matrices. Both the time reversed pionor defined in \eqref{eq:TimeReflectionSpinor} and the space reversed pinor defined in \eqref{eq:TimeReflectionSpinor} are part of the original pinor space $S$.

The two Dirac adjoints of a column {\it pinor} as defined in \eqref{eq:firstDiracAdjointSpinor} and \eqref{eq:secondDiracAdjointSpinor} are now isomorphic to {\it quaternionic} row vectors with $2^{(d-2)/2}$ entries. They differ essentially by a factor $\hat \gamma$. 

Column and row pinors can be multiplied with quaternions (column pinors from the right, row pinors from the left) to yield another such pinor. The spinor inner products defined in \eqref{eq:InnerProductSpinor1} and \eqref{eq:InnerProductSpinor2} yield an element of the quaternions $\mathbbm{H}$. 

Relativistic fermions described by these pinors correspond to {\it quaternionic fermions} and allow the definition of {\it symplectic Majorana fermions} with some additional structure (see eq.\ \eqref{eq:symplecticMajoranaSpinorCondition}).

\subsubsection{Cases with $r-(d-r)=1,5 \text{ mod }8$}\label{sec:pinor15}
Examples in up to five dimensions are $d=1+0$, $d=0+3$, $d=2+1$, $d=1+4$, $d=3+2$, $d=5+0$.

Here the real Clifford algebra $\mathcal{C}l(r,d-r,\mathbbm{R})$ is isomorphic to a {\it complex} matrix algebra $\text{Mat}(N, \mathbbm{C})$ with $N=2^{(d-1)/2}$ dimensions. Accordingly, column pinors are isomorphic to a column vector with $2^{(d-1)/2}$ {\it complex} entries. 

The Clifford structure map $\varsigma$ is given (up to an overall sign) by complex conjugation.  The ``volume element'' $\hat \gamma$ as defined in \eqref{eq:defhatgamma} and the elements for time reversal $\beta$ (defined in \eqref{eq:definitionBeta}) and space reversal $\alpha$ (defined in \eqref{eq:definitionBetaBar}) can all be represented by {\it complex} matrices. However, either time or space reversal also encompasses a complex conjugation $\varsigma$. Both the time reversed pionor defined in \eqref{eq:TimeReflectionSpinor} and the space reversed pinor defined in \eqref{eq:TimeReflectionSpinor} are part of the original pinor space $S$.

The two Dirac adjoints of a column {\it pinor} as defined in \eqref{eq:firstDiracAdjointSpinor} and \eqref{eq:secondDiracAdjointSpinor} are now isomorphic to {\it complex} row vectors with $2^{(d-1)/2}$ entries. They differ essentially by a complex conjugation. 

Column and row pinors can be multiplied with complex numbers to yield another such pinor. The spinor inner products defined in \eqref{eq:InnerProductSpinor1} and \eqref{eq:InnerProductSpinor2} yield both complex numbers.

Relativistic fermions described by these pinors correspond to {\it complex Dirac fermions}.

\subsubsection{Cases with $r-(d-r)=3 \text{ mod }8$}\label{sec:pinor3}
Examples in up to five dimensions are $d=3+0$, $d=0+5$, $d=4+1$. 

Here the real Clifford algebra $\mathcal{C}l(r,d-r,\mathbbm{R})$ is isomorphic to a {\it direct sum of quaternionic} matrix algebras $\text{Mat}(N/2, \mathbbm{H}) \oplus \text{Mat}(N/2, \mathbbm{H})$ with $N=2^{(d-1)/2}$. Column pinors can be either part of a minimal ideal, in which case they are isomorphic to a column vector that is non-zero in only one of the direct summands. In that case they have $2^{(d-3)/2}$ {\it quaternionic} entries corresponding to $2^{(d+1)/2}$ real dimensions. Alternatively, they can be part of a {\it quasi-minimal} ideal and are then isomorphic to to a direct sum of two column vectors wich has entries in both direct summands. In that case they together comprise $2^{(d-1)/2}$ {\it quaternionic} entries corresponding to $2^{(d+3)/2}$ real dimensions.

The Clifford structure map $\varsigma$ is given (up to an overall sign) by the interchange to the two direct summands.  The ``volume element'' $\hat \gamma$ as defined in \eqref{eq:defhatgamma} and the elements for time reversal $\beta$ (defined in \eqref{eq:definitionBeta}) and space reversal $\alpha$ (defined in \eqref{eq:definitionBetaBar}) can all be represented by {\it quaternionic} matrices in the direct sum of algebras. However, either time or space reversal also encompass an interchange of the two direct summands $\varsigma$. Accordingly, when the pinor space $S$ is a minimum ideal, the corresponding pinors are taken out of this space by either time reversal or space reversal and therefore break these symmetries in this sense. In contrast, when the pinor space is quasi-minimal, the time and space reversed pinors are part of the original pinor space.

The two Dirac adjoints of a column {\it pinor} as defined in \eqref{eq:firstDiracAdjointSpinor} and \eqref{eq:secondDiracAdjointSpinor} are in the minimal case isomorphic to {\it quaternionic} row vectors with $2^{(d-3)/2}$ entries. Because they differ by a power of the structure map $\varsigma$, they are in fact part of the two different direct summands. In the quasi-minimal case the two Dirac adjoints are isomorphic to {\it quaternionic} row vectors with $2^{(d-1)/2}$ entries. 

Column and row pinors can be multiplied with quaternions (column pinors from the right, row pinors from the left) to yield another such pinor. The spinor inner products defined in \eqref{eq:InnerProductSpinor1} and \eqref{eq:InnerProductSpinor2} yield an element of the quaternions $\mathbbm{H}$.

Relativistic fermions described by these pinors correspond to {\it minimal or quasi-minimal quaternionic fermions} and allow the definition of {\it minimal or quasi-minimal symplectic Majorana fermions} with some additional structure (see eq.\ \eqref{eq:symplecticMajoranaSpinorCondition}).

\subsubsection{Cases with $r-(d-r)=7 \text{ mod }8$}\label{sec:pinor7}
Examples in up to five dimensions are $d=0+1$, $d=1+2$, $d=2+3$. 

Here the real Clifford algebra $\mathcal{C}l(r,d-r,\mathbbm{R})$ is isomorphic to a {\it direct sum of real} matrix algebras $\text{Mat}(N/2, \mathbbm{R}) \oplus \text{Mat}(N/2, \mathbbm{R})$ with $N=2^{(d+1)/2}$. Column pinors can be either part of a minimal ideal, in which case they are isomorphic to a column vector that is non-zero in only one of the direct summands. In that case they have $2^{(d-1)/2}$ {\it real} entries. Alternatively, they can be part of a {\it quasi-minimal} ideal and are then isomorphic to to a direct sum of two column vectors wich has entries in both direct summands. In that case they together comprise $2^{(d+1)/2}$ {\it real} entries.

The Clifford structure map $\varsigma$ is given (up to an overall sign) by the interchange to the two direct summands.  The ``volume element'' $\hat \gamma$ as defined in \eqref{eq:defhatgamma} and the elements for time reversal $\beta$ (defined in \eqref{eq:definitionBeta}) and space reversal $\alpha$ (defined in \eqref{eq:definitionBetaBar}) can all be represented by {\it real} matrices in the direct sum of algebras. However, either time or space reversal also encompass an interchange of the two direct summands $\varsigma$. Accordingly, when the pinor space $S$ is a minimum ideal, the corresponding pinors are taken out of this space by either time reversal or space reversal and therefore break these symmetries in this sense. In contrast, when the pinor space is quasi-minimal, the time and space reversed pinors are part of the original pinor space.

The two Dirac adjoints of a column {\it pinor} as defined in \eqref{eq:firstDiracAdjointSpinor} and \eqref{eq:secondDiracAdjointSpinor} are in the minimal case isomorphic to {\it real} row vectors with $2^{(d-1)/2}$ entries. Because they differ by a power of the structure map $\varsigma$, they are in fact part of the two different direct summands. In the quasi-minimal case the two Dirac adjoints are isomorphic to {\it real} row vectors with $2^{(d+1)/2}$ entries. 

Column and row pinors can be multiplied with real numbers to yield another such pinor. The spinor inner products defined in \eqref{eq:InnerProductSpinor1} and \eqref{eq:InnerProductSpinor2} yield a real number.

Relativistic fermions described by these pinors correspond to {\it minimal or quasi-minimal Majorana fermions}.

\subsection{Proper spinors}
Let us now also discuss spinor spaces that feature non-trivial representations of the {\it even real} Clifford sub-algebra $\mathcal{C}l_+(r,d-r,\mathbbm{R})$. These spaces feature only a representation of the spin group $\text{Spin}(r,d-r,\mathbbm{R})$ (but not of the pin group), such that they it's elements are called {\it proper spinors}.

We take the spinor space $S$ to be a {\it minimal} or {\it quasi-minimal} left ideal $\mathcal{C}l_+(r,d-r,\mathbbm{R}) p$ where $p$ is a {\it minimal} or {\it quasi-minimal} idempotent (of the even sub-algebra), see the discussion in section \ref{sec:MatrixRepresentations}. 

In the following we go through the different cases where the real Clifford algebra is isomorphic to different matrix algebras. For this discussion it is useful to keep the classification of even sub-algebras in table \ref{tab:EvenCliffordAlgebraClassification} in mind.

\subsubsection{Cases with $r-(d-r)=2,6 \text{ mod } 8$}\label{sec:spinor26}
Examples in up to six dimensions are $d=0+2$, $d=2+0$, $d=1+3$ (Minkowski space), $d=3+1$, $d=0+6$, $d=2+4$, $d=4+2$, $d=6+0$.

Here the real, even Clifford sub-algebra $\mathcal{C}l_+(r,d-r,\mathbbm{R})$ is isomorphic to a real, full Clifford algebra $\mathcal{C}l(r^\prime,d^\prime-r^\prime,\mathbbm{R})$ with $d^\prime=d-1$ and $r^\prime-(d^\prime-r^\prime)=1,5 \text{ mod } 8$. The latter is isomorphic to a {\it complex} matrix algebra $\text{Mat}(N,\mathbbm{C})$ with $N=2^{(d-2)/2}$. 
Accordingly, column spinors as representations of the even sub-algebra are isomorphic to a column vector with $2^{(d-2)/2}$ {\it complex} entries corresponding to $2^{d/2}$ real dimensions. Interestingly, this real dimension is for $r-(d-r)=6 \text{ mod } 8$ as large as the corresponding pinor real dimension (see subsection \ref{sec:pinor06}) and for $r-(d-r)=2 \text{ mod } 8$ a factor $2$ smaller (see subsection \ref{sec:pinor24}).

Note that the structure map in the {\it full} Clifford algebra $\varsigma=\pm\hat \gamma$ commutes with all elements of the even sub-algebra $\mathcal{C}l_+(r,d-r,\mathbbm{R})$ and that $\hat\gamma^2=-\mathbbm{1}$. For irreducible representations it must be proportional to the unit matrix. One can understand $\hat \gamma$ as a complex structure and in fact there are two in-equivalent, complex conjugate representations where $\hat \gamma=+i\mathbbm{1}$ and $\hat \gamma=-i\mathbbm{1}$. The Clifford structure map $\varsigma$ of the {\it even} sub-algebra $\mathcal{C}l_+(r,d-r,\mathbbm{R})$ is a complex conjugation with respect to this complex structure.

The two Dirac adjoints of a column {\it spinor} as defined in \eqref{eq:firstDiracAdjointSpinor} and \eqref{eq:secondDiracAdjointSpinor} are now isomorphic to {\it complex} row vectors with $2^{(d-2)/2}$ entries. They differ essentially by a power of the structure map $\varsigma$, which is a complex conjugation. 

Column and row spinors can be multiplied with complex numbers to yield another such spinor. The spinor inner products defined in \eqref{eq:InnerProductSpinor1} and \eqref{eq:InnerProductSpinor2} yield both complex numbers. 

Relativistic fermions described by these spinors correspond to {\it complex Weyl fermions}.

\subsubsection{Cases with $r-(d-r)=4 \text{ mod } 8$}\label{sec:spinor4}
Examples in up to six dimensions are $d=0+4$ (Euclidean space), $d=4+0$, $d=1+5$, $d=5+1$. 

Here the real, even Clifford sub-algebra $\mathcal{C}l_+(r,d-r,\mathbbm{R})$ is isomorphic to a real, full Clifford algebra $\mathcal{C}l(r^\prime,d^\prime-r^\prime,\mathbbm{R})$ with $d^\prime=d-1$ and $r^\prime-(d^\prime-r^\prime)=3 \text{ mod } 8$. The latter is isomorphic to a {\it quaternionic direct sum} matrix algebra $\text{Mat}(N/2,\mathbbm{H}) \oplus \text{Mat}(N/2,\mathbbm{H})$ with $N=2^{(d-2)/2}$. 
Accordingly, column spinors as representations of the even sub-algebra are isomorphic to a direct sum of two column vectors with together $2^{(d-2)/2}$ {\it quaternionic} entries corresponding to $2^{(d+2)/2}$ real dimensions. This real dimension is as large as the one of the corresponding pinor representation (see subsection \ref{sec:pinor24}). On the other side, an irreducible spinor as part of a minimal ideal is non-zero in only one of these summands and has only $2^{(d-4)/2}$ {\it quaternionic} entries corresponding to $2^{d/2}$ real dimensions. This is then a factor $2$ smaller than the corresponding pinor representation.

Note that the structure map in the {\it full} Clifford algebra $\varsigma=\pm\hat \gamma$ commutes with all elements of the even sub-algebra $\mathcal{C}l_+(r,d-r,\mathbbm{R})$ and that $\hat\gamma^2=\mathbbm{1}$. For irreducible representations it must be proportional to the unit matrix. There are two in-equivalent representations where $\hat \gamma=+\mathbbm{1}$ and $\hat \gamma=-\mathbbm{1}$ corresponding to the two direct summands of the even sub-algebra $\mathcal{C}l_+(r,d-r,\mathbbm{R})$. The Clifford structure map $\varsigma$ of the {\it even} sub-algebra $\mathcal{C}l_+(r,d-r,\mathbbm{R})$ interchanges these two summands, $a\oplus b \to \pm b \oplus a$.

The two Dirac adjoints of a column {\it spinor} as defined in \eqref{eq:firstDiracAdjointSpinor} and \eqref{eq:secondDiracAdjointSpinor} are now isomorphic to {\it quaternionic} row vectors (in the irreducible case) or direct sums thereof. They differ essentially by a power of the structure map $\varsigma$, which interchanges the two direct summands.

Column and row spinors can be multiplied with quaternions (column spinors from the right, row spinors from the left) to yield another such spinor. The spinor inner products defined in \eqref{eq:InnerProductSpinor1} and \eqref{eq:InnerProductSpinor2} yield both an element of the quaternions $\mathbbm{H}$. 

Relativistic fermions described by these spinors might be called to {\it quaternionic Weyl fermions} and one may define {\it symplectic Majorana-Weyl fermions} with some additional structure (see eq.\ \eqref{eq:symplecticMajoranaSpinorCondition}).

\subsubsection{Cases with $r-(d-r)=0 \text{ mod } 8$}\label{sec:spinor0}
Examples in up to six dimensions are $d=1+1$, $d=2+2$, $d=3+3$.

Here the real, even Clifford sub-algebra $\mathcal{C}l_+(r,d-r,\mathbbm{R})$ is isomorphic to a real, full Clifford algebra $\mathcal{C}l(r^\prime,d^\prime-r^\prime,\mathbbm{R})$ with $d^\prime=d-1$ and $r^\prime-(d^\prime-r^\prime)=7 \text{ mod } 8$. The latter is isomorphic to a {\it real direct sum} matrix algebra $\text{Mat}(N/2,\mathbbm{R}) \oplus \text{Mat}(N/2,\mathbbm{R})$ with $N=2^{(d+1)/2}$. 
Accordingly, column spinors as representations of the even sub-algebra are isomorphic to a direct sum of two column vectors with together $2^{d/2}$ {\it real} entries. This real dimension is as large as the one of the corresponding pinor representation (see subsection \ref{sec:pinor06}). On the other side, an irreducible spinor as part of a minimal ideal is non-zero in only one of these summands and has only $2^{(d-2)/2}$ {\it real} entries. This is then a factor $2$ smaller than the corresponding pinor representation.

Note that the structure map in the {\it full} Clifford algebra $\varsigma=\pm\hat \gamma$ commutes with all elements of the even sub-algebra $\mathcal{C}l_+(r,d-r,\mathbbm{R})$ and that $\hat\gamma^2=\mathbbm{1}$. For irreducible representations it must be proportional to the unit matrix. There are two in-equivalent representations where $\hat \gamma=+\mathbbm{1}$ and $\hat \gamma=-\mathbbm{1}$ corresponding to the two direct summands of the even sub-algebra $\mathcal{C}l_+(r,d-r,\mathbbm{R})$. The Clifford structure map $\varsigma$ of the {\it even} sub-algebra $\mathcal{C}l_+(r,d-r,\mathbbm{R})$ interchanges these two summands, $a\oplus b \to \pm b \oplus a$.

The two Dirac adjoints of a column {\it spinor} as defined in \eqref{eq:firstDiracAdjointSpinor} and \eqref{eq:secondDiracAdjointSpinor} are now isomorphic to {\it real} row vectors (in the irreducible case) or direct sums thereof. They differ essentially by a power of the structure map $\varsigma$, which interchanges the two direct summands.

Column and row spinors can be multiplied with real numbers to yield another such spinor. The spinor inner products defined in \eqref{eq:InnerProductSpinor1} and \eqref{eq:InnerProductSpinor2} yield both a real number. 

Relativistic fermions described by these spinors are {\it Majorana-Weyl fermions}.

\subsubsection{Cases with $r-(d-r)=1,7 \text{ mod } 8$}\label{sec:spinor17}
Examples in up to five dimensions are $d=1+0$, $d=0+1$, $d=1+2$, $d=2+1$, $d=2+3$, $d=3+2$. 

Here the real, even Clifford sub-algebra $\mathcal{C}l_+(r,d-r,\mathbbm{R})$ is isomorphic to a real, full Clifford algebra $\mathcal{C}l(r^\prime,d^\prime-r^\prime,\mathbbm{R})$ with $d^\prime=d-1$ and $r^\prime-(d^\prime-r^\prime)=0, 6 \text{ mod } 8$. The latter is isomorphic to a {\it real} matrix algebra $\text{Mat}(N,\mathbbm{R})$ with $N=2^{(d-1)/2}$. 
Accordingly, column spinors as representations of the even sub-algebra are isomorphic to column vectors with $2^{(d-1)/2}$ {\it real} entries. This real dimension is a factor $2$ smaller than the  one of the corresponding pinor representation (see subsection \ref{sec:pinor15}) for $r-(d-r)=1 \text{ mod } 8$ and as large as an irreducible pinor representation for $r-(d-r)=7 \text{ mod } 8$. The transition from the pinor to the spinor representation reduces for $r-(d-r)=1 \text{ mod }8$ a complex to a real representation and for $r-(d-r)=7 \text{ mod } 8$ a direct sum of two real representations to an irreducible real representation.

The Clifford structure map $\varsigma$ of the {\it even} sub-algebra $\mathcal{C}l_+(r,d-r,\mathbbm{R})$ is now given by the product of all generators and is actually an element of the even Clifford algebra itself. The two Dirac adjoints of a column {\it spinor} as defined in \eqref{eq:firstDiracAdjointSpinor} and \eqref{eq:secondDiracAdjointSpinor} are now isomorphic to {\it real} row vectors. They differ essentially by a power of the structure map.

Column and row spinors can be multiplied with real numbers to yield another such spinor. The spinor inner products defined in \eqref{eq:InnerProductSpinor1} and \eqref{eq:InnerProductSpinor2} yield both a real number. 

In odd dimensions there is now chiral symmetry and therefore no Weyl fermions.  Relativistic fermions described by these spinors may be called {\it Majorana fermions}.

\subsubsection{Cases with $r-(d-r)=3,5 \text{ mod } 8$}\label{sec:spinor35}
Examples in up to five dimensions are $d=0+3$, $d=3+0$, $d=0+5$, $d=1+4$, $d=4+1$, $d=5+0$. 

Here the real, even Clifford sub-algebra $\mathcal{C}l_+(r,d-r,\mathbbm{R})$ is isomorphic to a real, full Clifford algebra $\mathcal{C}l(r^\prime,d^\prime-r^\prime,\mathbbm{R})$ with $d^\prime=d-1$ and $r^\prime-(d^\prime-r^\prime)=2, 4 \text{ mod } 8$. The latter is isomorphic to a {\it quaternionic} matrix algebra $\text{Mat}(N,\mathbbm{R})$ with $N=2^{(d-3)/2}$.

Accordingly, column spinors as representations of the even sub-algebra are isomorphic to column vectors with $2^{(d-3)/2}$ {\it quaternionic} entries corresponding to $2^{(d+1)/2}$ real dimensions. This real dimension is as large as the one of the corresponding pinor representations (see subsections \ref{sec:pinor15} and \ref{sec:pinor3}). In other words, in this case there if no reduction in the number of real degrees of freedom occurring in the transition from the pinor to the spinor representation.

The Clifford structure map $\varsigma$ of the {\it even} sub-algebra $\mathcal{C}l_+(r,d-r,\mathbbm{R})$ is now given by the product of all generators and is actually an element of the even Clifford algebra itself. The two Dirac adjoints of a column {\it spinor} as defined in \eqref{eq:firstDiracAdjointSpinor} and \eqref{eq:secondDiracAdjointSpinor} are now isomorphic to {\it quaternionic} row vectors. They differ essentially by a power of the structure map.

Column and row spinors can be multiplied with quaternions (column spinors from the right, row spinors from the left) to yield another such spinor. The spinor inner products defined in \eqref{eq:InnerProductSpinor1} and \eqref{eq:InnerProductSpinor2} yield both an element of the quaternions $\mathbbm{H}$. 

In odd dimensions there is now chiral symmetry and therefore no Weyl fermions.  Relativistic fermions described by these spinors may be called {\it quaternionic fermions} and allow the construction of {\it symplectic Majorana fermions} with some additional structure (see eq.\ \eqref{eq:symplecticMajoranaSpinorCondition}).

\section{Conclusions}

We have discussed here real Clifford algebras as well as their representations in terms of {\it pinors} and {\it proper spinors} from an algebraic perspective with the goal to put the description of relativistic fermions in various dimensions on a common basis. Independent of a concrete matrix representation, {\it pinor} spaces have been introduced as minimal or quasi-minimal left and right ideals within the {\it full} Clifford algebra $\mathcal{C}l(r,d-r,\mathbbm{R})$. They feature in particular a non-trivial representation of the {\it pin} group. In a similar way, {\it proper spinor} spaces have been introduced as minimal or quasi-minimal left and right ideals within the {\it even} sub-algebra $\mathcal{C}l_+(r,d-r,\mathbbm{R})$. They feature naturally a representation of the {\it spin} group. 

An advantage of the present approach is that it provides a unified treatment of Clifford algebras and corresponding spinors in all combinations of time dimensions $r$ and space dimensions $d-r$. An important role is played by the Clifford structure map $\varsigma$ as defined in eq.\ \eqref{eq:StructureMapCliffordAlgebra}. For example, the action of the pin group on an arbitrary element of the Clifford algebra has been formulated in eq.\ \eqref{eq:actionPinGroupCliffordAlgebraElement} in terms of powers of the structure map. This form was in fact crucial to define the action of the pin group on pinors, equation \eqref{eq:pinGroupOnSpinors} such that the transformed pinor stays within the corresponding quasi-minimal left ideal.

Another important ingredient of our construction was the definition of an ``hermitian conjugation'' as an involution within the Clifford algebra, see \eqref{eq:defHermitianConjugateGenerator}. This allowed to define column spinors and row spinors based on the same minimal (or quasi-minimal) canonical idempotent $p$ and also to define two Dirac adjoints in eqs.\ \eqref{eq:firstDiracAdjointSpinor} and \eqref{eq:secondDiracAdjointSpinor} in a rather general way. The latter differ essentially by a power of the structure map and depending on $d$ and $r$, one of these ``Dirac adjoints'' is traditionally known as the ``Majorana adjoint''. It is also nice to see how the Clifford conjugate $\mathsf{C}(a)$ and the reverse $\mathsf{R}(a)$ of a pin group element $a$ appear naturally in the general transformation behavior of the Dirac adjoints under the pin group, equation \eqref{eq:actionPinGroupDiracAdjoints}. Together with the analysis of the pin group in section \ref{sec:PinAndSpinGroups}, one can now easily determine how spinors and their inner products transform under various discrete space-time symmetries.

An interesting general lesson from studying real Clifford algebras is that spinors spaces are either real, complex or quaternionic vector spaces and that also inner products between spinors yield either real, complex or quaternionic numbers. This is interesting because such products directly enter the Lagrangian or action that defines the physical theory as well as effective actions, correlation functions and so on. Of course, another important ingredient to construct possible theories in the form of actions is the anti-commuting Grasmann nature of fermionic fields, which has not been discussed here.

Beyond the setup of the current study, it is also interesting to investigate the analytic continuation between spaces with equal number of dimension $d$ but different signature $r-(d-r)$ \cite{Frohlich:1974zs, Mehta:1986mi, vanNieuwenhuizen:1996tv, Belitsky:2000ii, Wetterich:2010ni}. In fact, this needs as a first step a {\it complexification} of the entire Clifford algebra. The complex case is actually interesting by itself and will be subject of a future study. Besides the problem of analytic continuation we are also motivated by possible extension of the spinor helicity formalism to other dimensions (for reviews about the latter see refs.\ \cite{Dixon:1996wi, Elvang:2013cua, Henn:2014yza}, for a recent extension to massive theories and arbitrary spin ref.\ \cite{Arkani-Hamed:2017jhn}).

\subsection*{Acknowledgements}
The author thanks C. Wetterich for discussions. This work is supported by Deutsche Forschungsgemeinschaft (DFG) under EXC-2181/1-390900948 (the Heidelberg STRUCTURES Excellence Cluster), SFB 1225 (ISOQUANT) as well as BE 2795/4-1.

\end{document}